\documentclass[ALICE,manyauthors]{cernphprep}
\usepackage[comma,square,numbers,sort&compress]{natbib}

\usepackage{lineno}
\modulolinenumbers[5]
\usepackage{xspace}
\usepackage{hyperref}
\usepackage{color}
\usepackage[T1]{fontenc}
\usepackage{orcidlink}

\begin{document}
%

\newcommand{\pp}           {pp\xspace}
\newcommand{\ppbar}        {\mbox{$\mathrm {p\overline{p}}$}\xspace}
\newcommand{\XeXe}         {\mbox{Xe--Xe}\xspace}
\newcommand{\PbPb}         {\mbox{Pb--Pb}\xspace}
\newcommand{\pA}           {\mbox{pA}\xspace}
\newcommand{\pPb}          {\mbox{p--Pb}\xspace}
\newcommand{\AuAu}         {\mbox{Au--Au}\xspace}
\newcommand{\dAu}          {\mbox{d--Au}\xspace}

\newcommand{\s}            {\ensuremath{\sqrt{s}}\xspace}
\newcommand{\snn}          {\ensuremath{\sqrt{s_{\mathrm{NN}}}}\xspace}
\newcommand{\pt}           {\ensuremath{p_{\rm T}}\xspace}
\newcommand{\meanpt}       {$\langle p_{\mathrm{T}}\rangle$\xspace}
\newcommand{\ycms}         {\ensuremath{y_{\rm CMS}}\xspace}
\newcommand{\ylab}         {\ensuremath{y_{\rm lab}}\xspace}
\newcommand{\etarange}[1]  {\mbox{$\left | \eta \right |~<~#1$}}
\newcommand{\yrange}[1]    {\mbox{$\left | y \right |~<~#1$}}
\newcommand{\dndy}         {\ensuremath{\mathrm{d}N_\mathrm{ch}/\mathrm{d}y}\xspace}
\newcommand{\dndeta}       {\ensuremath{\mathrm{d}N_\mathrm{ch}/\mathrm{d}\eta}\xspace}
\newcommand{\avdndeta}     {\ensuremath{\langle\dndeta\rangle}\xspace}
\newcommand{\dNdy}         {\ensuremath{\mathrm{d}N_\mathrm{ch}/\mathrm{d}y}\xspace}
\newcommand{\Npart}        {\ensuremath{N_\mathrm{part}}\xspace}
\newcommand{\Ncoll}        {\ensuremath{N_\mathrm{coll}}\xspace}
\newcommand{\dEdx}         {\ensuremath{\textrm{d}E/\textrm{d}x}\xspace}
\newcommand{\RpPb}         {\ensuremath{R_{\rm pPb}}\xspace}

\newcommand{\nineH}        {$\sqrt{s}~=~0.9$~Te\kern-.1emV\xspace}
\newcommand{\seven}        {$\sqrt{s}~=~7$~Te\kern-.1emV\xspace}
\newcommand{\twoH}         {$\sqrt{s}~=~0.2$~Te\kern-.1emV\xspace}
\newcommand{\twosevensix}  {$\sqrt{s}~=~2.76$~Te\kern-.1emV\xspace}
\newcommand{\five}         {$\sqrt{s}~=~5.02$~Te\kern-.1emV\xspace}
\newcommand{\twosevensixnn}{$\sqrt{s_{\mathrm{NN}}}~=~2.76$~Te\kern-.1emV\xspace}
\newcommand{\fivenn}       {$\sqrt{s_{\mathrm{NN}}}~=~5.02$~Te\kern-.1emV\xspace}
\newcommand{\LT}           {L{\'e}vy-Tsallis\xspace}
\newcommand{\GeVc}         {Ge\kern-.1emV/$c$\xspace}
\newcommand{\MeVc}         {Me\kern-.1emV/$c$\xspace}
\newcommand{\TeV}          {Te\kern-.1emV\xspace}
\newcommand{\GeV}          {Ge\kern-.1emV\xspace}
\newcommand{\MeV}          {Me\kern-.1emV\xspace}
\newcommand{\GeVmass}      {Ge\kern-.1emV/$c^2$\xspace}
\newcommand{\MeVmass}      {Me\kern-.1emV/$c^2$\xspace}
\newcommand{\lumi}         {\ensuremath{\mathcal{L}}\xspace}

\newcommand{\ITS}          {\rm{ITS}\xspace}
\newcommand{\TOF}          {\rm{TOF}\xspace}
\newcommand{\ZDC}          {\rm{ZDC}\xspace}
\newcommand{\ZDCs}         {\rm{ZDCs}\xspace}
\newcommand{\ZNA}          {\rm{ZNA}\xspace}
\newcommand{\ZNC}          {\rm{ZNC}\xspace}
\newcommand{\SPD}          {\rm{SPD}\xspace}
\newcommand{\SDD}          {\rm{SDD}\xspace}
\newcommand{\SSD}          {\rm{SSD}\xspace}
\newcommand{\TPC}          {\rm{TPC}\xspace}
\newcommand{\TRD}          {\rm{TRD}\xspace}
\newcommand{\VZERO}        {\rm{V0}\xspace}
\newcommand{\VZEROA}       {\rm{V0A}\xspace}
\newcommand{\VZEROC}       {\rm{V0C}\xspace}
\newcommand{\Vdecay} 	   {\ensuremath{V^{0}}\xspace}

\newcommand{\ee}           {\ensuremath{e^{+}e^{-}}} 
\newcommand{\pip}          {\ensuremath{\pi^{+}}\xspace}
\newcommand{\pim}          {\ensuremath{\pi^{-}}\xspace}
\newcommand{\kap}          {\ensuremath{\rm{K}^{+}}\xspace}
\newcommand{\kam}          {\ensuremath{\rm{K}^{-}}\xspace}
\newcommand{\pbar}         {\ensuremath{\rm\overline{p}}\xspace}
\newcommand{\kzero}        {\ensuremath{{\rm K}^{0}_{\rm{S}}}\xspace}
\newcommand{\lmb}          {\ensuremath{\Lambda}\xspace}
\newcommand{\almb}         {\ensuremath{\overline{\Lambda}}\xspace}
\newcommand{\Om}           {\ensuremath{\Omega^-}\xspace}
\newcommand{\Mo}           {\ensuremath{\overline{\Omega}^+}\xspace}
\newcommand{\X}            {\ensuremath{\Xi^-}\xspace}
\newcommand{\Ix}           {\ensuremath{\overline{\Xi}^+}\xspace}
\newcommand{\Xis}          {\ensuremath{\Xi^{\pm}}\xspace}
\newcommand{\Oms}          {\ensuremath{\Omega^{\pm}}\xspace}
\newcommand{\degree}       {\ensuremath{^{\rm o}}\xspace}

\newcommand{\Jpsi}         {\ensuremath{{\rm J}/\psi}\xspace}
\newcommand{\Psip}         {\ensuremath{\psi^\prime}\xspace}
\newcommand{\Noon}         {\textbf{n$\mathbf{_O^O}$n}}
\newcommand{\WgPb}{\ensuremath{W_{\gamma\, \mathrm{Pb, n}}}\xspace}
\newcommand{\gPb}{\gamma\,\mathrm{Pb}}
\newcommand{\ptdimuon}     {\ensuremath{p^{\mu\mu}_{\rm T}}\xspace}
\newcommand{\mdimuon}     {\ensuremath{m_{\mu\mu}}\xspace}
\newcommand{\pttwo}        {\ensuremath{p^{2}_{\rm T}}\xspace}
\newcommand{\mant}         {\ensuremath{|t|}\xspace}
\newcommand{\ggll}         {\ensuremath{\gamma\gamma\to l^+ l^-}\xspace}

\newcommand{\BR}           {\ensuremath{\rm BR}\xspace}
\newcommand{\axe}          {\ensuremath{\rm{Acc\times\epsilon}}\xspace}
\newcommand{\fC}           {\ensuremath{f_{\rm C}}\xspace}
\newcommand{\fD}           {\ensuremath{f_{\rm D}}\xspace}

\newcommand{\rojo}[1]{{\color{red}{#1}}}
\newcommand{\azul}[1]{{\color{blue}{#1}}}

\begin{titlepage}
\PHyear{2023}       
\PHnumber{100}      
\PHdate{30 May}  

\title{
Energy dependence of coherent photonuclear production of $\mathbf{\Jpsi}$ mesons in ultra-peripheral \PbPb collisions at $\mathbf{\sqrt{\textit{s}_{\mathrm{NN}}}~=~5.02}$~Te\kern-.1emV}
\ShortTitle{Energy dependence of coherent photonuclear production of $\Jpsi$ mesons}   

\Collaboration{ALICE Collaboration\thanks{See Appendix~\ref{app:collab} for the list of collaboration members}}
\ShortAuthor{ALICE Collaboration} 

\begin{abstract}
The cross section for coherent photonuclear production of $\Jpsi$ is presented as a function of 
the electromagnetic  dissociation (EMD) of Pb. The measurement is performed with the ALICE detector in  ultra-peripheral  \PbPb collisions  at a centre-of-mass energy per nucleon pair of \fivenn.
Cross sections are presented in five different $\Jpsi$ rapidity ranges within $|y|<4$, with the $\Jpsi$ reconstructed via its dilepton decay channels. 
In some events the $\Jpsi$ is not accompanied by EMD, while other events do produce neutrons from EMD at beam rapidities either in one or the other beam direction, or in both. The cross sections in a given rapidity range and for different configurations of neutrons from EMD allow for the extraction of the energy  dependence of  this process in the range $17 < \WgPb <920$ \GeV, where $\WgPb$ is the centre-of-mass energy per nucleon of the $\gPb$ system. This range corresponds to a Bjorken-$x$ interval spanning about three orders of magnitude: $ 1.1\times10^{-5}<x<3.3\times 10^{-2}$. In addition to the ultra-peripheral and photonuclear cross sections, the nuclear suppression factor is obtained. These measurements point to a strong depletion of the gluon distribution in Pb nuclei over a broad, previously unexplored, energy range.
These results, together with previous ALICE measurements, provide unprecedented information to probe quantum chromodynamics at high energies.
\end{abstract}

\end{titlepage}

\setcounter{page}{2} 


\section{Introduction}

One of the main research topics in quantum chromodynamics (QCD) today is the study of the hadronic structure when probed at high energies, corresponding to low values of the fraction of the hadron momentum carried
by the colliding parton (Bjorken-$x$). The  gluon distribution inside the proton has been observed to increase steeply at low values of $x$~\cite{H1:2015ubc}.  At some point, this growth must stop to preserve unitarity. In QCD this is achieved by a dynamic equilibrium of  gluon splitting and annihilation processes. This regime of the gluon distribution is known as saturation, see Ref.~\cite{Morreale:2021pnn} for a recent review. In a nucleus with $A$ nucleons, the parton distributions would naively be $A$ times those in a single nucleon, but modifications, known as nuclear shadowing, are observed at small $x$~\cite{Armesto:2006ph}. 
 Similar considerations to those for a single nucleon regarding unitarity imply that saturation is expected to set in for large nuclei at lower energies (higher $x$ values)  than in  protons, with this behaviour scaling roughly as $A^{1/3}$~\cite{McLerran:1993ni}.

Diffractive production of $\Jpsi$ vector mesons off nuclear targets is a powerful tool 
to study the energy evolution of the structure of heavy nuclei. The interaction can involve the full nucleus or only one nucleon; these cases are called coherent and incoherent production, respectively. The coherent process has a large experimental cross section, it is very sensitive to the  gluon structure of hadrons, and  it can be
described within perturbative QCD owing to the large $\Jpsi$ mass, which provides a hard scale  to justify the use of perturbative techniques.
At the LHC, the coherent production of $\Jpsi$ can be measured in ultra-peripheral collisions (UPCs) where the incoming Pb nuclei pass each other at impact parameters larger than the sum of their radii, such that the interaction involves  photons from the strong electromagnetic field of the incoming ions~\cite{Baltz:2007kq,Contreras:2015dqa,Klein:2019qfb,Klein:2020fmr}.

Previous measurements of this process at the LHC were performed at different centre-of-mass energies per nucleon pair (\snn) and different rapidities  of the $\Jpsi$. Using data from  LHC Run 1, where the Pb nuclei  collided at \twosevensixnn, the ALICE Collaboration measured the coherent photoproduction of $\Jpsi$s in UPCs at forward rapidity~\cite{Abelev:2012ba} and midrapidity~\cite{Abbas:2013oua}, while the CMS Collaboration provided a cross section for this process at an intermediate rapidity range~\cite{Khachatryan:2016qhq}. For LHC Run 2, the energy of \PbPb collisions was raised to \fivenn and new measurements of this process were performed by the ALICE Collaboration at mid~\cite{ALICE:2021gpt,ALICE:2021tyx} and forward rapidity~\cite{Acharya:2019vlb} as well as  by the LHCb Collaboration at forward rapidity~\cite{LHCb:2021bfl}.

The importance of a wide experimental rapidity coverage is that the rapidity  of the $\Jpsi$ in this process is related to the centre-of-mass energy per nucleon in the  $\gPb$ system by $(\WgPb)^2 = m \sqrt{s_{\rm NN}} \exp(-y)$, where $m$ is the mass of the $\Jpsi$ and $y$ its rapidity in the laboratory frame measured with respect to the direction of the incoming Pb nucleus. (Natural units are used in all equations.)  At the LHC, either of the two incoming Pb ions can be the source of the photon and, in this circumstance, the cross section for the coherent photoproduction of $\Jpsi$ in UPCs as a function of rapidity has two components~\cite{Klein:1999qj}
\begin{equation}
\frac{{\rm d}\sigma_{\rm PbPb}}{{\rm d}y} = n_\gamma(y,\{b\})\sigma_{\gPb}(y)+n_\gamma(-y,\{b\})\sigma_{\gPb}(-y),
\label{eq:upc}
\end{equation}
where $\sigma_{\gPb}(y)$ is the photonuclear cross section for the coherent production of  a $\Jpsi$ at rapidity $y$, and $n_\gamma(y)$ is the
photon flux which, in the equivalent photon approximation~\cite{Baltz:2007kq}, quantifies the number of photons with energy $k = (m/2)\exp(-y)$. The notation $\{b\}$ signifies that the flux is obtained by integrating over a  range on impact parameter $b$.

A study of the rapidity dependence of $\sigma_{\gPb}$ was performed in Ref.~\cite{Guzey:2013xba} using ALICE Run~1 data at \twosevensixnn.  The analysis is based on  two facts: (1) at $y=0$ both contributions in Eq.~(\ref{eq:upc}) are equal, so that  knowledge of the photon flux $n_\gamma(y=0)$ yields  $\sigma_{\gPb}$ at the corresponding $\WgPb=92$~\GeV; (2) at the largest rapidities accessible to ALICE, the first term in the right-hand side of Eq.~(\ref{eq:upc})
contributes only about 5\% 
so that the UPC cross section is dominated by the second term in Eq.~(\ref{eq:upc}) corresponding to interactions with low-energy photons at $\WgPb=20$ \GeV for the ALICE data used in the analysis.
The extracted cross sections are discussed in Sec.~\ref{sec:gPb}.

In order to extract the full energy dependence of $\sigma_{\gPb}$ from the UPC cross section, at least  two measurements at the same rapidity but with different photon fluxes are needed. Up to now, there are two proposals on how to achieve this. Both utilise the fact that the photon flux also depends on the impact-parameter range where the $\gPb$ interaction takes place. One proposal, presented in Ref.~\cite{Contreras:2016pkc}, makes use of the coherent production of $\Jpsi$ measured in {\em peripheral} collisions originally reported by the ALICE Collaboration~\cite{ALICE:2015mzu} and later on confirmed by the STAR~\cite{STAR:2019yox}, ALICE~\cite{ALICE:2022zso}, and LHCb~\cite{LHCb:2021hoq} Collaborations. 
Applying this approach to ALICE data at \twosevensixnn for the coherent photonuclear production of $\Jpsi$ measured in peripheral collisions and in UPCs, in Ref.~\cite{Contreras:2016pkc} the cross sections for $\sigma_{\gPb}$ at three values of $\WgPb$: 18 \GeV, 92 \GeV, and 470 \GeV are obtained.  The extracted cross sections are discussed in Sec.~\ref{sec:gPb}.
The other proposal, presented in Refs.~\cite{Baltz:2002pp,Guzey:2013jaa}, utilises the fact that the electromagnetic fields of the incoming nuclei are so strong that there is a sizeable probability of a second photon exchange between the colliding nuclei, which may result in the electromagnetic dissociation (EMD) of at least one of the interacting nuclei~\cite{Pshenichnov:2011zz}. The presence of neutrons from  EMD of one or both nuclei, which can be determined using zero-degree calorimeters, can be used to tag specific ranges of the impact parameter. 
This is so because high energy photons are emitted at smaller impact parameters than low energy photons and in order to induce the dissociation of a nucleus a photon needs a minimum energy of the order of 10 MeV. This means that events where EMD has occurred select a photon flux with $\{b\}$ covering smaller impact parameters than events without EMD. 
In this way, the measurement of coherent $\Jpsi$ photoproduction in UPCs with no, single, or mutual EMD can be used to disentangle the two  different $\sigma_{\gPb}$ contributions in 
Eq.~(\ref{eq:upc})~\cite{Guzey:2013jaa}.
The tagging of single and mutual EMD
 has been successfully tested; first, in the measurement of coherent $\rho^0$ photoproduction at midrapidity by the ALICE Collaboration~\cite{ALICE:2015nbw,ALICE:2020ugp,ALICE:2021jnv}, and later  in the measurement of $\gamma\gamma\to\mu^+\mu^-$ and $\gamma\gamma\to e^+e^-$ in UPCs  by the CMS and  ATLAS Collaborations~\cite{CMS:2020skx,ATLAS:2020epq,ATLAS:2022srr}. These ATLAS and CMS measurements have been successfully described by the newest version of the SuperChic Monte Carlo generator~\cite{Harland-Lang:2023ohq}.
 More recently, the CMS Collaboration submitted results on the coherent photoproduction of $\Jpsi$ accompanied by EMD in a  rapidity range complementary to the one explored in this analysis~\cite{CMS:2023snh}.

In this article, the cross section for  the coherent photoproduction of $\Jpsi$ accompanied by nuclear EMD is presented (Sec.~\ref{sec:UPC}). The measurement is carried out in five rapidity regions covering the intervals $|y|<0.8$ and $2.5<|y|<4.0$. These  measurements  are used to extract the photonuclear cross section $\sigma_{\gPb}$ in the range $17 < \WgPb <920$ \GeV, which corresponds to a Bjorken-$x$ in the range $1.1\times10^{-5}<x<3.3\times 10^{-2}$, where $x=m^2/\WgPb^2$ (Sec.~\ref{sec:gPb}). In addition, the nuclear suppression factor is obtained in this kinematic region (Sec.~\ref{sec:nsp}). The measurements are compared to theoretical models covering a wide spectrum of approaches going from models assuming no nuclear dynamics, to state-of-the-art computations based on perturbative QCD.

\section{Theoretical models
\label{sec:Models}}
Many different models provide predictions for the coherent photoproduction of $\Jpsi$ in UPCs; for example,  those presented in 
Refs.~\cite{Klein:1999qj,Guzey:2016piu,Lappi:2013am,Cepila:2017nef,Mantysaari:2017dwh,Bendova:2020hbb,Goncalves:2020vdp,Eskola:2022vpi}.
Predictions for this process when accompanied by EMD of the incoming nuclei exist for only  a few of the models, which are discussed in Sec.~\ref{sec:CrossSection}.
All the models are based on the computation of the two elements shown in 
Eq.~(\ref{eq:upc}):  the photon flux $n_\gamma(y)$ and  the photonuclear cross section $\sigma_{\gPb}(y)$. There is also an interference term~\cite{Klein:1999gv}, but its effect, when the cross section is integrated over the transverse momentum of the $\Jpsi$  as for  the results presented here,  can be  neglected.

\subsection{The photon flux\label{sec:Flux}}
The predictions and the experimental results discussed below utilise photon fluxes based on the approach of the STARlight~\cite{Klein:1999qj,Klein:2016yzr} and \Noon\ models~\cite{Broz:2019kpl}. 
There are two steps to compute the photon fluxes: obtaining the total flux, and computing the fractions of the flux that are assigned to the different EMD classes presented in Sec.~\ref{sec:EMD}. 

The total flux as implemented in both  STARlight and \Noon\ 
is computed in the semi-classical approximation (for details see, e.g. Ref.~\cite{Baltz:2007kq}). In this approach, the form factor of the Pb ion is modelled with a Woods--Saxon distribution, while the coherence condition is supplemented by the requirement of no hadronic interactions, as obtained with a Poissonian model based on the nuclear overlap function and the total nucleon--nucleon cross section.

Both  STARlight~\cite{Baltz:2002pp} and  \Noon~\cite{Broz:2019kpl} compute the fractions of the total fluxes for each EMD scenario  based on  photoproduction data measured at lower energies and extrapolations to LHC energies. 
There are two relevant differences between STARlight and \Noon.
 First, STARlight uses the 
Lorentz-line parameterisation of the giant-dipole resonance data described in Ref.~\cite{Veyssiere:1970ztg}, while \Noon\ uses the data directly; the numerical difference between them is negligible. Second, \Noon\ uses photonuclear Pb data in the nucleon resonance region~\cite{Bianchi:1995vb},  while STARlight uses data from photon--nucleon interactions~\cite{Armstrong:1971ns,Armstrong:1972sa} in this region. 
Both models provide similar fluxes apart from the most forward rapidity region. In this kinematic range, the fluxes differ by up to 20\%.
 As \Noon\ uses experimental data on $\gPb$ collisions, the fluxes from this model are used in Sec.~\ref{sec:gPb} to extract the photonuclear cross section.

\subsection{The photonuclear cross section\label{sec:CrossSection}
}
The photonuclear cross section can be computed using various theoretical approaches. The impulse approximation (IA) assumes that the nuclear scattering is given by the superposition of the scattering on the individual nucleons~\cite{Chew:1952fca}.
In the context of the coherent production of $\Jpsi$, a nuclear suppression factor can be defined using IA, and the associated cross section $\sigma^{\rm IA}_{\gPb}$. This factor quantifies the difference between the nucleus being a set of independent nucleons and a real nucleus:
\begin{equation}
S_{\rm Pb}(\WgPb) = \sqrt{\frac{\sigma_{\gPb}}{\sigma^{\rm IA}_{\gPb}}}.
\label{eq:nsf}
\end{equation}
The square root in Eq.~(\ref{eq:nsf}) is motivated by the fact that the diffractive photoproduction of $\Jpsi$ is proportional to the square of the gluon distribution of the target within the leading log approximation of QCD~\cite{Ryskin:1992ui}. 

The model for the photonuclear cross section by Klein and Nystrand~\cite{Klein:1999qj}, implemented in STARlight, is based on the following steps.  A parameterisation of HERA data on the exclusive forward production of $\Jpsi$ is converted, using the vector dominance model (VDM)~\cite{Bauer:1977iq}, into the forward cross section for $\Jpsi + {\rm p} \to \Jpsi + {\rm p}$; using the optical theorem, this cross section yields the total $\Jpsi+{\rm p}$ cross section, which is introduced into   a classical Glauber prescription to produce the total cross section for $\Jpsi +{\rm Pb}$. Finally, the optical theorem and VDM are used again to obtain  the forward $\sigma_{\gPb}$. The nuclear form factor is used to obtain the total $\sigma_{\gPb}$.

The model by Guzey, Kryshen, and Zhalov~\cite{Guzey:2016piu} is based on the leading logarithmic approximation of perturbative QCD~\cite{Ryskin:1992ui} for the  exclusive production of $\Jpsi$ at zero momentum transfer for $\gamma\,{\rm p}$ collisions. This cross section is scaled to the nuclear case using the square of the ratio of the gluon distribution in the Pb nucleus to the gluon distribution in the proton scaled by the Pb mass number.
The computation is performed for  two cases. The first one is based on the EPS09-LO parameterisation of nuclear parton density functions~\cite{Eskola:2009uj}. The second one relies on the leading twist approximation (LTA) of gluon shadowing~\cite{Frankfurt:2011cs}. This model includes the nuclear form factor computed with a Woods--Saxon prescription. The photon fluxes  needed to compute the UPC cross section are obtained from the flux fractions given by STARlight. The theoretical uncertainties explored in this model originate in the spread of predictions from the nuclear parton distribution functions (PDFs) for the EPS09-LO case, and in the uncertainty on the parameters of LTA obtained by fits to HERA diffractive data. For the LTA case, the uncertainty on the predicted cross sections reaches up to 30\%, while for EPS09-LO it can be as large as a factor of 2.
The authors of the LTA computations provided the upper and lower limits of their predictions. The average of these numbers is depicted in the figures shown in Sec.~\ref{sec:Results}. These figures also show the predictions for the EPS09-based model, using the central value of the EPS09 parameterisation. 

The model by Bendova et al.~\cite{Bendova:2020hbb} is based on the solution of the impact-parameter dependent  Balitsky--Kovchegov (b-BK) equation, as discussed in Ref.~\cite{Bendova:2019psy} and references therein. There are two models presented in Ref.~\cite{Bendova:2020hbb}. 
One uses the b-BK equation to evolve the amplitude for the interaction of a colour dipole with a proton towards higher energies, and then uses this amplitude to compute the $\gamma\,\mathrm{p}$ cross section at the given energy followed by the  application of the Glauber--Gribov approach~\cite{Gribov:1968jf}, in order to obtain the photonuclear cross section. 
 The second model, shown in the figures below with the notation b-BK-A, starts with a nuclear initial condition for the b-BK equation whose solutions at higher energies  can then be directly used to obtain the photonuclear cross section  without the need of a Glauber--Gribov prescription.
The photon fluxes needed to compute the UPC cross section are given by the \Noon\ model. There are some theoretical uncertainties associated with this type of model~\cite{Cepila:2019skb}.
The main two uncertainties are
related to the use of the Glauber--Gribov approach
 instead of a nuclear initial condition, and to  the approach used to compute the $\Jpsi$ wave function. The first uncertainty changes the cross section for coherent production of $\Jpsi$  up to 30\%~\cite{Bendova:2020hbb}, while the $\Jpsi$ wave function produces an uncertainty up to about 20\%~\cite{Mantysaari:2017dwh,Krelina:2018hmt}. This model is valid only at small Bjorken-$x$, so the cross section in UPCs for $|y|$ larger than about three cannot be predicted as these rapidities are dominated by contributions with Bjorken-$x$ larger than 0.01.

The model by Cepila et al.~\cite{Cepila:2017nef} is based on the colour-dipole approach to QCD, including gluon saturation effects, framed within the Good--Walker formalism for diffraction~\cite{Good:1960ba,Miettinen:1978jb,Klein:2023zlf}. 
In this model,  hadrons are constituted by hot spots with the hadronic structure fluctuating event by event. For a recent review, see Ref.~\cite{Mantysaari:2020axf}. This type of model describes  HERA data~\cite{Mantysaari:2016ykx,Cepila:2016uku}. 
In the  model by Cepila et al. the number of hot spots  increases as Bjorken-$x$ decreases and the transition from proton to nuclear targets is based on the Glauber--Gribov prescription.
The theoretical uncertainties coming from the wave function of the $\Jpsi$ are the same as described above. Other uncertainties related to the fluctuation of the colour fields were explored in Ref.~\cite{Mantysaari:2016ykx} and found to be small with respect to the current precision of the experimental data. 
The uncertainty on the modelling of the nuclear case was explored in Ref.~\cite{Cepila:2017nef} by comparing the Glauber--Gribov prescription with results based on a geometric-scaling approach. 
Differences of up to 30\% were found, with recent data~\cite{ALICE:2021gpt} clearly preferring the Glauber--Gribov prescription, which is shown in the figures below with the notation GG-HS. Also in this case, the photon fluxes needed to compute the UPC cross section are given by the \Noon\ model.

\section{Experimental set-up}

The results presented here are based on a data sample collected with the ALICE detector in 2018, when the LHC provided collisions of Pb nuclei at \fivenn. The experimental signature of the events of interest for this analysis consists of a pair of leptons, from the $\Jpsi$ decay, the potential presence of neutrons emitted at beam rapidities by EMD, and no other signal above the noise threshold recorded in the detector. The two tracks produced by the leptons  are measured  with the central barrel detectors (dimuon and dielectron decay channels of the $\Jpsi$) to obtain the results for $|y|<0.8$, discussed in Sec.~\ref{sec:CB}, or with the muon spectrometer (dimuon  channel only)  to obtain the results for  $2.5<|y|<4.0$, described in Sec~\ref{sec:MS}. Forward detectors located in the A and C sides of the experiment\footnote{The A and C nomenclature is used LHC wide and refers to the direction of flight of the beams in the accelerator as being anti- or clockwise when the LHC is seen from the top.} are used to record the neutrons and to veto other activity; they are introduced in Sec.~\ref{sec:FD}. The triggers used in this analysis, and the associated luminosity, are presented in  Sec.~\ref{sec:TL}. The full description of the ALICE detector and its performance can be found in Refs.~\cite{Aamodt:2008zz,Abelev:2014ffa}.

\subsection{Central barrel detectors
\label{sec:CB}}

Three central barrel detectors, the Inner Tracking System (ITS), the Time Projection Chamber (TPC), and the Time-of-Flight (TOF) were used to record data for this analysis. These detectors  are surrounded by a large solenoid magnet producing a magnetic field of $B = 0.5$ T. 
Their common pseudorapidity acceptance is $|\eta|<0.9$.

The ITS~\cite{Aamodt:2010aa} consists of six cylindrical layers of silicon detectors. The innermost layer is at a radius of 3.9 cm with respect to the beam axis, while the outermost layer is at 43 cm. The two layers closest to the beam form the Silicon Pixel Detector (SPD) and cover the range in pseudorapidity  $|\eta|<1.4$. 
The SPD is a fine granularity detector with about 10 million pixels. It serves as a tracking device and can also be used to issue triggers. Surrounding the SPD there are two layers of silicon drift chambers and  then two  layers of silicon microstrips. These four outer  layers of the ITS are used in this analysis exclusively for tracking. 

The TPC~\cite{Alme:2010ke} is a five metre  long cylindrical chamber separated into two drift volumes by a 100 kV central electrode. The two end-plates are 250 cm away from the central electrode along the beam direction; they are instrumented with multi-wire proportional chambers that are readout by about 560\,000 pads allowing for high precision tracking in the transverse plane. The longitudinal coordinate is given by the drift time of ionisation electrons in the TPC electric field. For each individual track, the TPC provides up to 159 track points, which also provide  energy-loss measurements that are used for particle identification (PID). In the momentum range of the tracks considered in this analysis (from 1 to 2 \GeVc) the PID from the TPC allows for a clean separation of electrons from muons. The TPC covers the range $|\eta|<0.9$.

The TOF detector consists of a barrel of multi-gap resistive plate chambers that provide a  high precision timing for tracks traversing TOF~\cite{ALICE:2016ovj}.
It surrounds the TPC and has a pseudorapidity coverage of $|\eta|<0.9$. The TOF readout channels are arranged into 18 azimuth sectors that can provide topological trigger decisions~\cite{Akindinov:2009zzc}.

\subsection{The muon spectrometer
\label{sec:MS}}

The muon spectrometer, located in the C side of the experiment, covers the pseudorapidity interval $-4<\eta<-2.5$.
The composition of the muon spectrometer, when seen from the nominal interaction point (IP), is as follows. 
First, there is a ten hadronic interaction-length absorber---made of carbon, concrete, and steel---with the task of filtering out hadrons produced in the collisions. The absorber is followed by 
five tracking stations, each made of  two planes of cathode pad chambers. The third station is inside a dipole magnet producing  a 3 T\,m integrated magnetic field. The next element is an iron wall with a thickness of 7.2 hadronic interaction lengths, which is followed by the muon trigger system consisting of two stations, each instrumented  with two layers of resistive plate chambers.  In addition,  a conical absorber made of tungsten, lead, and steel  surrounds the beam pipe at small polar angles (less than $2^\circ$) with the mission of shielding the spectrometer from secondary particles.
Muon tracks detected in the trigger stations are used by the trigger and matched offline to the  tracks reconstructed in the five tracking stations. The trigger system provides single-muon and dimuon triggers for tracks above a programmable transverse-momentum threshold. 
For the  2018 data used in this analysis the threshold was set to 1 \GeVc. The trigger efficiency for tracks measured with both the trigger and the tracking chambers increases with transverse momentum and it is approximately 50\% at 1 \GeVc.

\subsection{Forward detectors
\label{sec:FD}}

The  zero-degree neutron calorimeters, ZNA and ZNC, are two 8.7 interaction-length calorimeters made of a tungsten alloy with embedded quartz fibres~\cite{Puddu2007,Oppedisano2009}. They are located $\pm112.5$ m, respectively, from the IP along the beam direction. They detect neutral particles produced at a pseudorapidity $|\eta| > 8.8$. Each calorimeter is segmented into four towers. Half the optical fibres, which are uniformly distributed in the calorimeter, are read out by four photomultipliers (PMTs) and the other half are read out by a single fifth photomultiplier common to all towers. 
The signals collected from the PMTs are used to determine the energy deposition.
The relative energy resolution for one neutron is about 20\%~\cite{ALICE:2012aa}.  In addition, each calorimeter provides timing information obtained with a TDC (time-to-digital converter). The detectors can also provide a trigger signal.

Two systems, V0~\cite{Abbas:2013taa} and AD~\cite{Broz:2020ejr}, are used to veto other activity in the events. Both V0 and AD are based on plastic scintillators and wave-length shifters; the light is captured using  PMTs.  Each system has two counters that are placed at both sides of the IP along the beam direction. 
The V0 counters have 32 scintillator tiles each. They cover the pseudorapidity ranges  $2.8 < \eta < 5.1$ (V0A) and $-3.7 < \eta < -1.7$ (V0C) and are located at 340 cm and 90 cm from the interaction point, respectively. 
The AD system consists of two arrays, each of 8 scintillator modules. The arrays are arranged in two layers of four modules. The AD arrays cover the pseudorapidity ranges $4.7<\eta<6.3$ (ADA) and $-6.9<\eta<-4.9$ (ADC), and are located 17 m and 19.5 m from the IP, respectively. Both V0 and AD have a timing resolution well below 1 ns, and both can be used to veto hadronic interactions at the trigger level.
 
\subsection{Triggers and luminosity
\label{sec:TL}}

The analysis is based on two triggers: one to select candidate events with the two leptons from the decay of the \Jpsi detected by the central barrel detectors (denoted as CBtrig below) and the other to select candidate events where the muons are measured by the muon spectrometer (denoted as MStrig below). They use trigger inputs from AD, V0, TOF, SPD and the muon trigger system. 

AD and V0 provide triggers based on the timing of the signal. Two time windows  are defined: one to trigger by events compatible with an interaction at the IP (beam--beam window) and another compatible with interactions happening behind one of the two counters of each system (beam--gas window).
In the triggers described below, the requested input is the logical negation of a trigger in the beam--beam window. For the V0 the following nomenclature is used: notVBA (notVBC) for the veto of activity in V0A (V0C). The corresponding triggers for AD are notUBA and notUBC.

The SPD is read out by 400 (800) chips in the inner (outer) layer with each of the readout chips providing a trigger signal if at least one of its pixels is fired. When projected into the transverse plane, the chips are arranged in 20 (40) azimuth regions in the inner (outer) layer allowing for a topological selection of events at the trigger level. In particular, a trigger element, denoted by STG below, requires at least two pairs of chips where each pair has a trigger signal in the inner and in the outer layer in the same azimuth region  and the pairs are back-to-back in azimuth.

A similar trigger, called OMU, is based on the trigger signals from TOF where between 2 and 6 signals from TOF are required, such that at least two of them are back-to-back in azimuth.

Using these elements CBtrig is the logical AND of: notVBA, notVBC, notUBA, notUBC, STG, and OMU. MStrig is given by the logical AND of MUL and notVBA, where MUL stands for a dimuon with the tracks having opposite electric charge and each of them is above the transverse-momentum threshold of 1 \GeVc as measured by  the trigger chambers of the muon spectrometer.

The integrated luminosity ($L_{\rm \, int}$) of the samples selected by the central barrel and the muon spectrometer triggers just described is determined using reference cross sections measured in van der Meer scans~\cite{ALICE:2022xir} and amounts  to $233\pm 7$ $\mu$b$^{-1}$ and $533\pm13$  $\mu$b$^{-1}$, respectively.

\section{Data samples}
\subsection{Event selection with the central barrel detectors
\label{sec:CBsel}}
The event selection is the same as that of Ref.~\cite{ALICE:2021gpt}. Events are kept for further analysis if:
\begin{itemize}
\item CBtrig is fired.
\item There is a reconstructed primary vertex, determined using at least two reconstructed tracks, and having a position within 15 cm of either side of the IP along the beam direction. 
\item There are, in the central barrel, exactly two good reconstructed tracks of opposite electric charge. Good reconstructed tracks are made of signals from the ITS and the TPC. Each track has to cross at least 70 (out of a maximum of 159) TPC pad rows and it also has to include signals from both SPD layers. Each track must have a distance of closest approach to the primary vertex, in the direction along the beam line, of less than 2 cm.
\item The transverse momentum of the pair fulfils $p_{\rm T}< 0.2$ \GeVc, in order to select a data sample enriched with coherently produced events.
\item Events have to pass the offline selection using the reconstructed information from V0 and AD.
 The offline selection in these detectors is more precise than vetoes at the trigger level, because it relies on larger time windows than the trigger electronics and on a more refined algorithm to quantify the signal.
\end{itemize}

The PID capabilities of the TPC are used to determine the mass to be associated with a track according to the proximity of the energy lost by ionisation to that expected by an electron or muon hypothesis.
The selected  events are distributed into two rapidity intervals, $|y|<0.2$ and $0.2<|y|<0.8$. They are the input for the midrapidity analysis described below.

\subsection{Event selection with the muon spectrometer
\label{sec:MSsel}}
The event selection follows closely that used in Ref.~\cite{Acharya:2019vlb}; the  difference being that in 2015 the trigger included, in addition to a veto in V0A as in 2018,  vetoes in ADA, ADC, and V0C. The 2018 data sample, which has less vetoes, is used
for this analysis. 
As the data sample is smaller in this analysis than for the results
presented in Ref.~\cite{Acharya:2019vlb}, the integrated luminosity is correspondingly
reduced. Note also, that after the publication of Ref.~\cite{Acharya:2019vlb}, the
luminosity determination was better understood~\cite{ALICE:2022xir} which reflects in a
substantially smaller uncertainty for the luminosity in the present
work.
Events are kept for further analysis if:
\begin{itemize}
\item MStrig is fired.
\item There are, in the muon spectrometer, exactly two good reconstructed tracks of opposite electric charge. Good reconstructed tracks have a pseudorapidity $-4<\eta<-2.5$; their radial position at the exit of the absorber  lies within 17.5 cm and 89.5 cm to ensure that they pass 
through the homogeneous region of the absorber; the information of the trigger chambers  matches that from the tracking chambers; the track momentum multiplied by the distance of closest approach of the track to the interaction point is below a set threshold to remove beam-induced background. 
\item The  four-momentum of the track pair,  constructed using the muon mass, has to have a rapidity in the range $2.5<|y|<4$. The transverse momentum of the pair is less than 0.25 \GeVc.
\item The event passes an offline veto, which is applied using the reconstructed information from V0A. At most two tiles in V0C have a signal in the beam--beam window, in order to allow for a maximum of two muons crossing this counter. 
\end{itemize}

The selected events are distributed into three rapidity intervals, $2.5<|y|<3.0$, $3.0<|y|<3.5$, and $3.5<|y|<4.0$. They are the input for the forward rapidity analysis described below.

\subsection{Event classification using ZNA and ZNC
\label{sec:EMD}}
The selected events  are  classified into three neutron classes depending on the presence of signals in the neutron zero-degree calorimeters. The presence of a neutron is determined using the timing capabilities of the calorimeter.
If the TDC registers a signal that has an energy over a threshold around 500 GeV, then the event is tagged as having at least one neutron emitted near beam rapidities. The following classes can be formed:
\begin{itemize}
\item No neutrons are registered either in ZNA or in ZNC. This class is denoted as 0n0n below.
\item At least one neutron is observed in one of the calorimeters, but not in the other. There are two cases:  ($i$) ZNA detects at least one neutron and ZNC shows no neutron activity, or ($ii$) ZNC detects neutron(s) and ZNA does not. They are denoted  as 0nXn or Xn0n, respectively. For the data sample at midrapidity, both cases are combined in one class: 0nXn+Xn0n. For the data sample at forward rapidity, the 0nXn sample has a large contamination from incoherent $\Jpsi$ production~\cite{Guzey:2013jaa}, so only the class Xn0n is considered for further analysis.
\item Both ZNA and ZNC detect at least one  neutron. This case is denoted as XnXn.
\end{itemize}

\subsection{Monte Carlo samples}

The STARlight Monte Carlo (MC)~\cite{Klein:2016yzr} (version: r299)  is used to generate event samples for the following five processes in \PbPb UPC: coherent and incoherent production of both  $\Jpsi \to l^+ l^-$ and $\Psip\to\Jpsi+X$ as well as $\ggll$, where $l$ denotes a lepton. 
The generated particles are propagated through a model of the ALICE experimental set-up implemented in GEANT 3.21~\cite{Brun:1082634}. The simulation matches the time evolution of the detector conditions during the data-taking period. The simulated data sets are passed through the same analysis chain as the real data, which allows for using the coherent MC sample to compute the acceptance and efficiency of the detector for signal, and all samples to be used to determine the background.

\section{Analysis procedure
\label{sec:Ana}}

The cross section for  coherent $\Jpsi$ photoproduction in UPC for a given neutron class and  rapidity interval is given by
\begin{equation}
\frac{{\rm d}\sigma_{\rm PbPb}}{{\rm d}y} = \frac{N_{\Jpsi}}{({\rm A}\times\epsilon) \times {\rm BR}(\Jpsi\to l^+l^-)\times L_{\rm \, int}\times \Delta y},
\label{eq:xs}
\end{equation} 
where $N_{\Jpsi}$ represents the $\Jpsi$ yield, $({\rm A}\times\epsilon)$ takes into account the acceptance and  efficiency of the detector, ${\rm BR}(\Jpsi\to l^+l^-)$ is the branching ratio, $L_{\rm \, int}$ stands for the integrated luminosity of the corresponding data sample, and $\Delta y$ is the width of the rapidity interval.

\subsection{Yield extraction
\label{sec:YE}}

\begin{figure}[!t]
\centering
\includegraphics[width=0.48\textwidth]{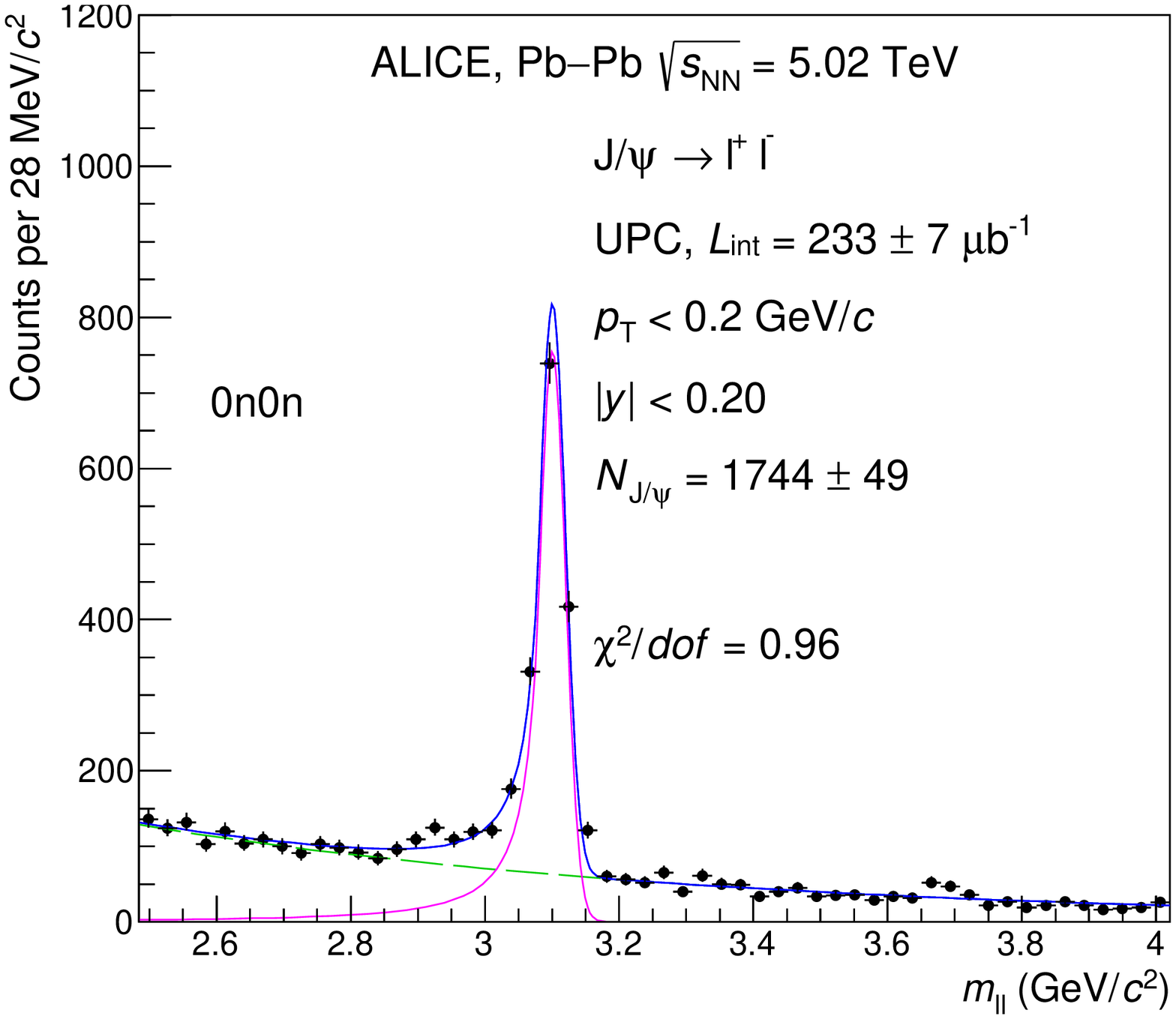}
\includegraphics[width=0.48\textwidth]{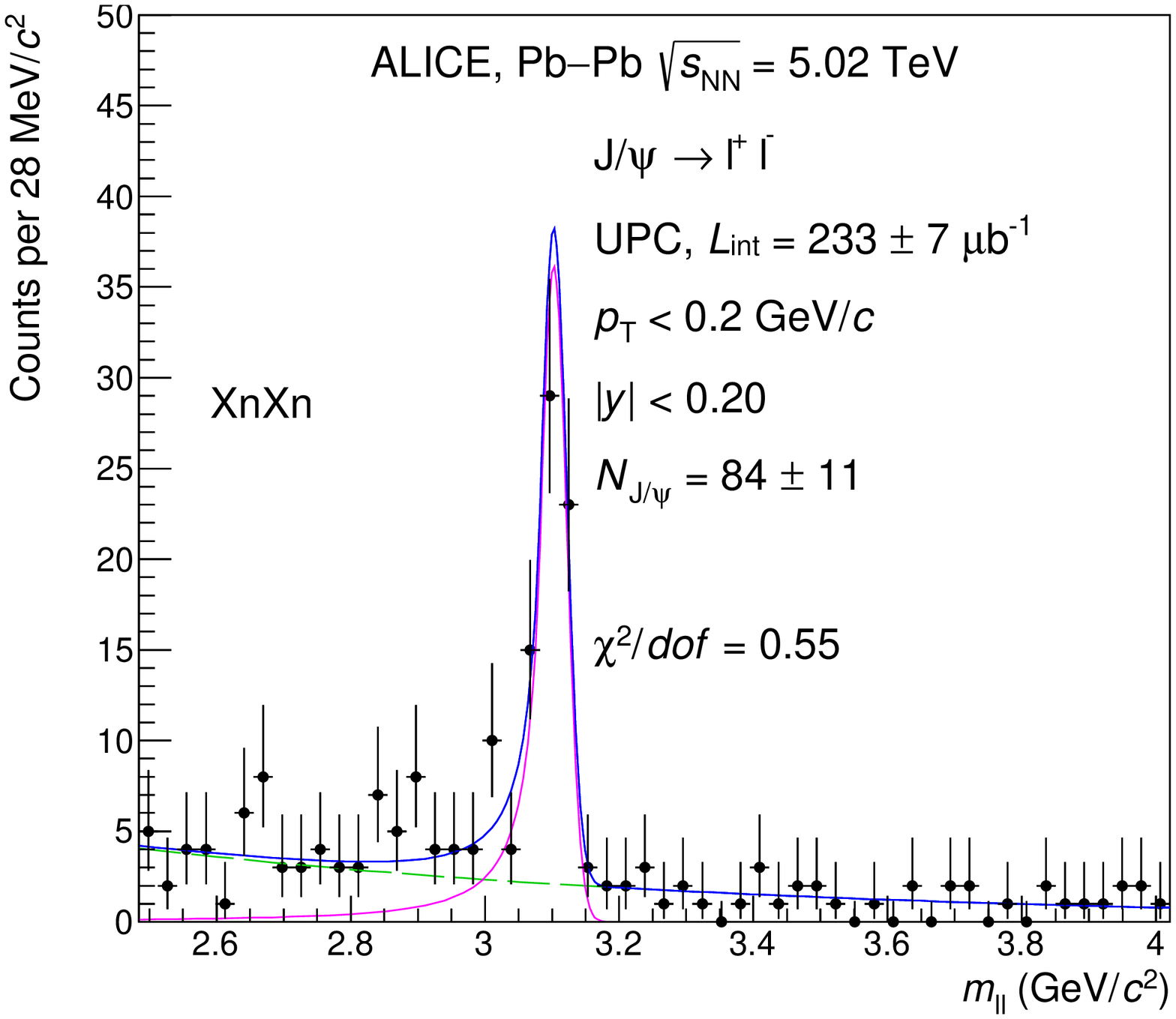}
\includegraphics[width=0.48\textwidth]{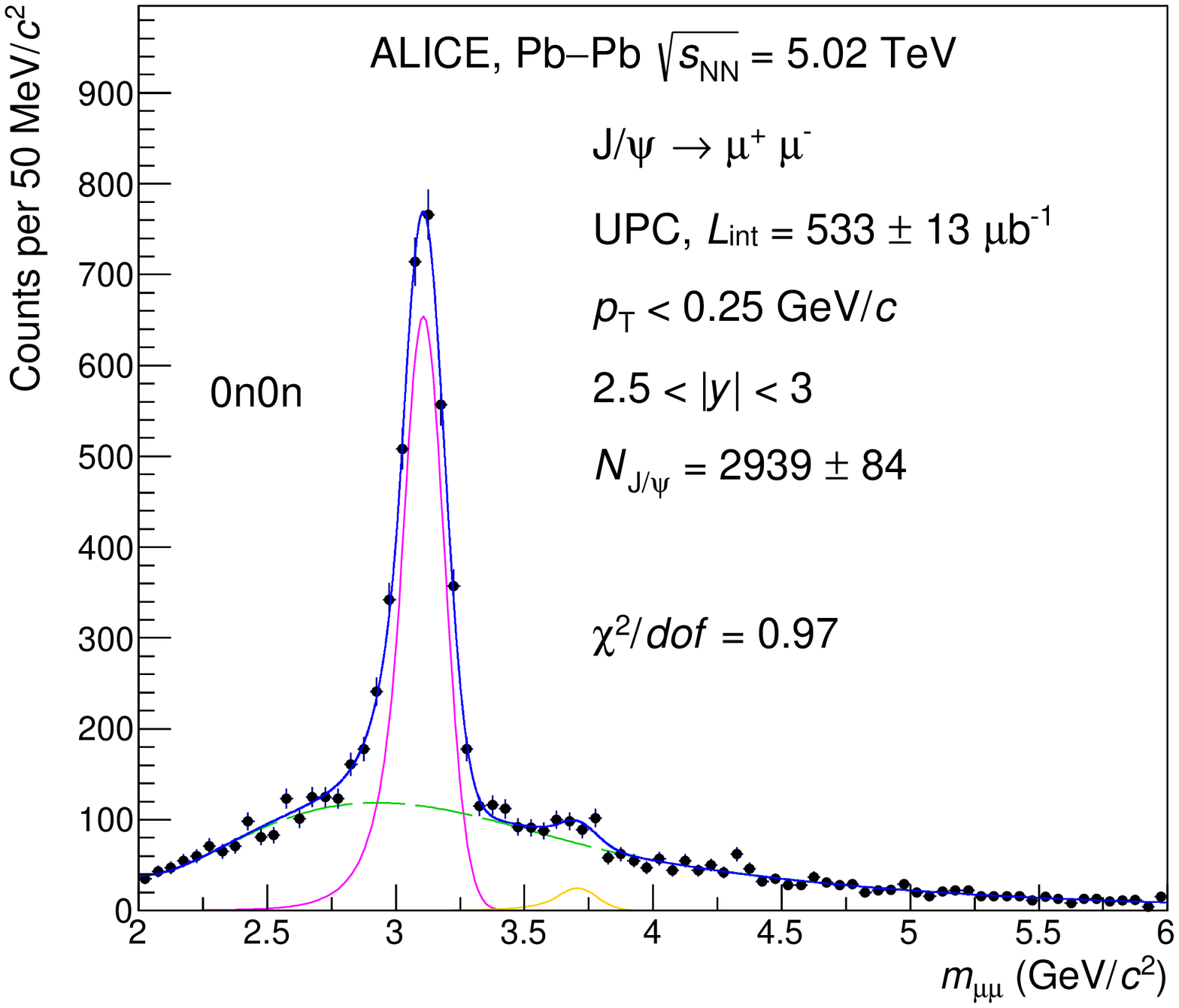}
\includegraphics[width=0.48\textwidth]{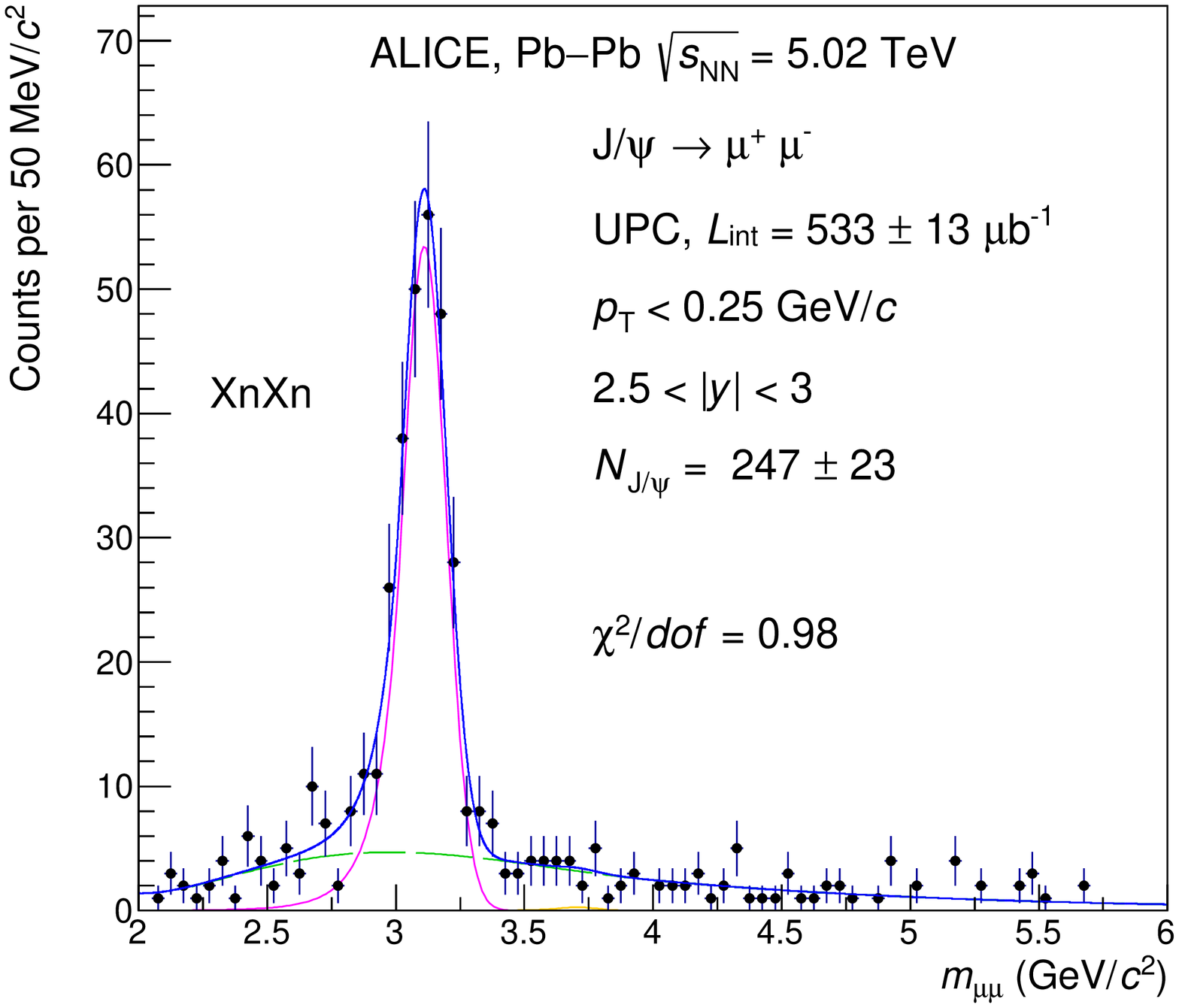}
\caption{
Invariant-mass distributions for events in the 0n0n (left) and XnXn (right) neutron classes measured at mid (top) and forward rapidity (bottom). The solid black markers represent data, the vertical line through each of them is the associated statistical uncertainty. The blue lines depict the fit models, described in the text, which are composed of a signal  (shown in magenta) and a background (shown in green) contribution. In the lower left plot the contribution from $\Psip$ is clearly visible (shown in yellow).}
\label{fig:mass}
\end{figure}

In order to make the best use of the limited amount of data at midrapidity,
specially for the XnXn neutron class, the data samples for the two decay
channels ($\Jpsi \to e^+e^-$ and $\Jpsi \to \mu^+\mu^-$) are joined
and analysed together. Note that in our previous analysis~\cite{ALICE:2021gpt},
it was demonstrated that both decay channels produce compatible cross sections for
coherent $\Jpsi$ production, justifying our decision to join the
samples.

The extraction of the $\Jpsi$ yield involves three steps: a fit to the invariant mass distribution of the lepton pairs; the subtraction of contributions from $\Psip$ feed-down and from incoherent $\Jpsi$ production; and a correction to account for the migration across the different neutron classes.

The invariant mass distribution is modelled with two Crystal Ball~\cite{Oreglia:1980cs} functions, to describe the $\Jpsi$ and $\Psip$ signals, and a term to take into account the background, mostly from two-photon production of dilepton pairs. For the analysis at midrapidity, this background is represented by an exponential distribution, while for the analysis at forward rapidity it is parameterised using a fourth-order polynomial that turns smoothly into an exponential tail for masses larger than 4~\GeVmass.
The tail parameters of the Crystal Ball functions are fixed to the values found by fits to the MC samples. 
The mass difference between the $\Psip$ and the $\Jpsi$ is fixed according to the values from Ref.~\cite{ParticleDataGroup:2022pth}. 
The width parameter for the $\Psip$ is fixed to the width parameter of the $\Jpsi$ multiplied by the ratio of the widths of the $\Psip$ to the
$\Jpsi$ obtained from fits to the MC samples. The $\Jpsi$ pole mass and width are left free. At forward rapidity, the polynomial parameters for the background are fixed to the values from a fit to the MC samples of the $\gamma\gamma\to\mu^+\mu^-$ process. The slope of the exponential is left free.
The main output of the fit is the number of $\Jpsi$ candidates ($N_{\rm fit}$) and its associated uncertainty. Examples of the fit to the invariant mass distribution for the mid and forward rapidity analyses for two different neutron classes are shown in Fig.~\ref{fig:mass}.

The number of candidates from the fit contains contributions from coherent and incoherent processes as well as from feed-down from $\Psip$. The number of coherent candidates is given by:
\begin{equation}
N_{\rm coh} = \frac{N_{\rm fit}}{1+f_{\rm I}+f_{\rm D}},
\label{eq:Ncoh}
\end{equation}
where the fractions $f_{\rm I} = N_{\rm incoh}/N_{\rm coh}$ and $f_{\rm D} = N_{\textrm {feed-down}}/N_{\rm coh}$ correct for the number of $\Jpsi$ coming from the incoherent process ($N_{\rm incoh}$) and from decays of $\Psip$ ($N_{\textrm{feed-down}}$). The fraction $f_{\rm D}$ is obtained directly from data as explained in Ref.~\cite{ALICE:2021gpt} for the midrapidity and in Ref.~\cite{Acharya:2019vlb} for the forward rapidity analysis. The values found are $f_{\rm D}=0.039\pm0.006$ and $f_{\rm D}=0.055\pm0.010$, respectively.

\begin{figure}[!t]
\centering
\includegraphics[width=0.48\textwidth]{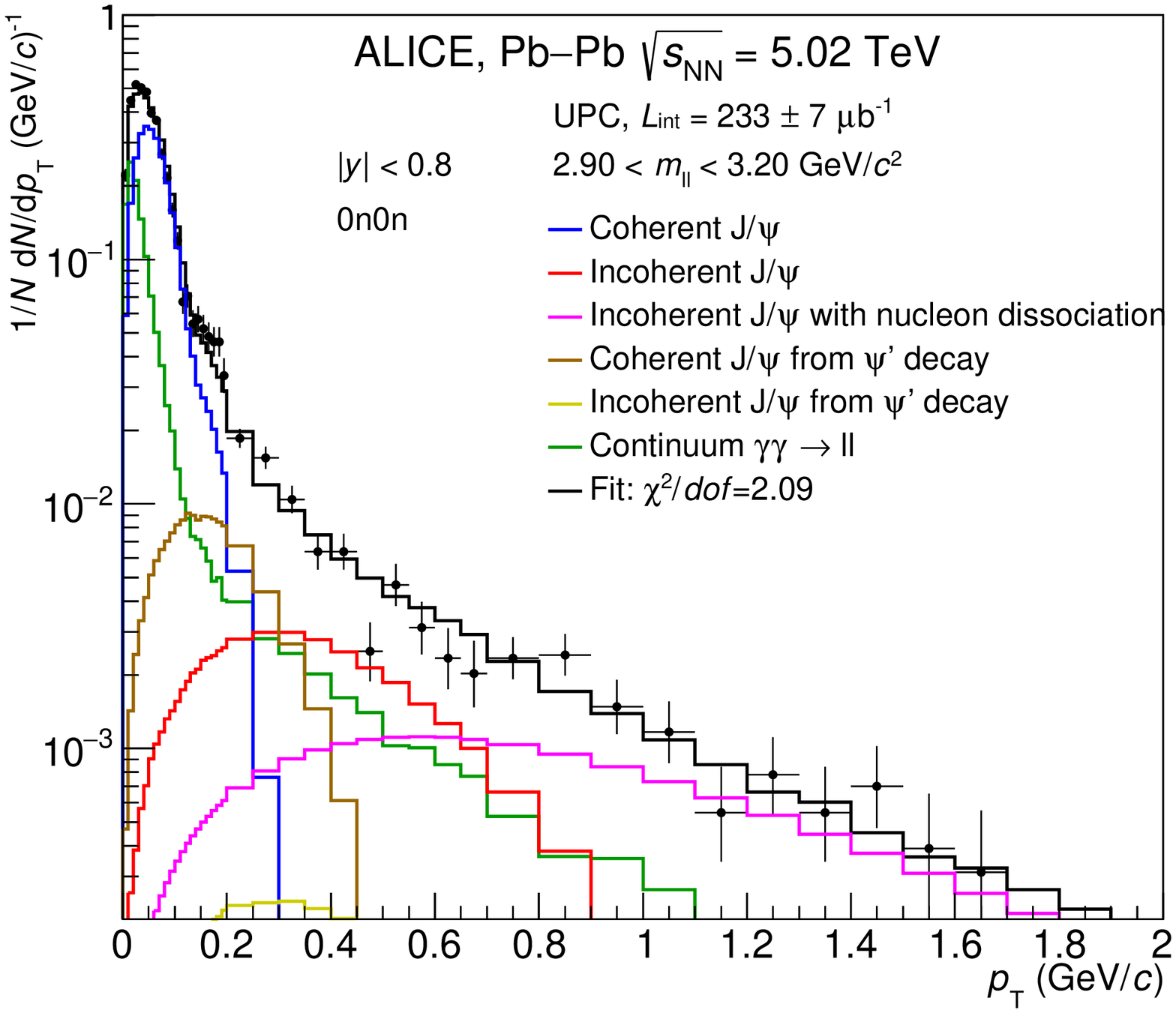}
\includegraphics[width=0.48\textwidth]{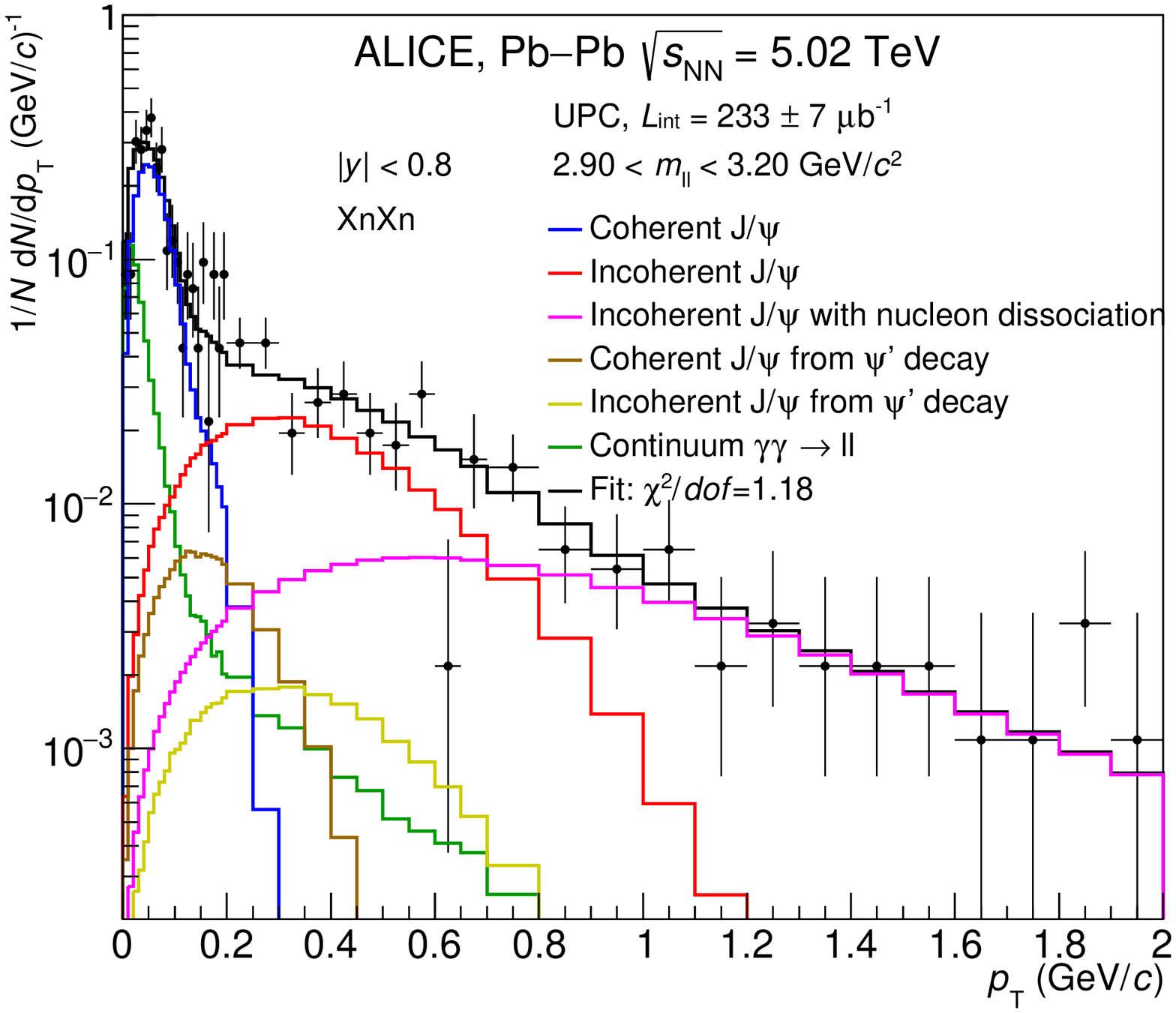}
\includegraphics[width=0.48\textwidth]{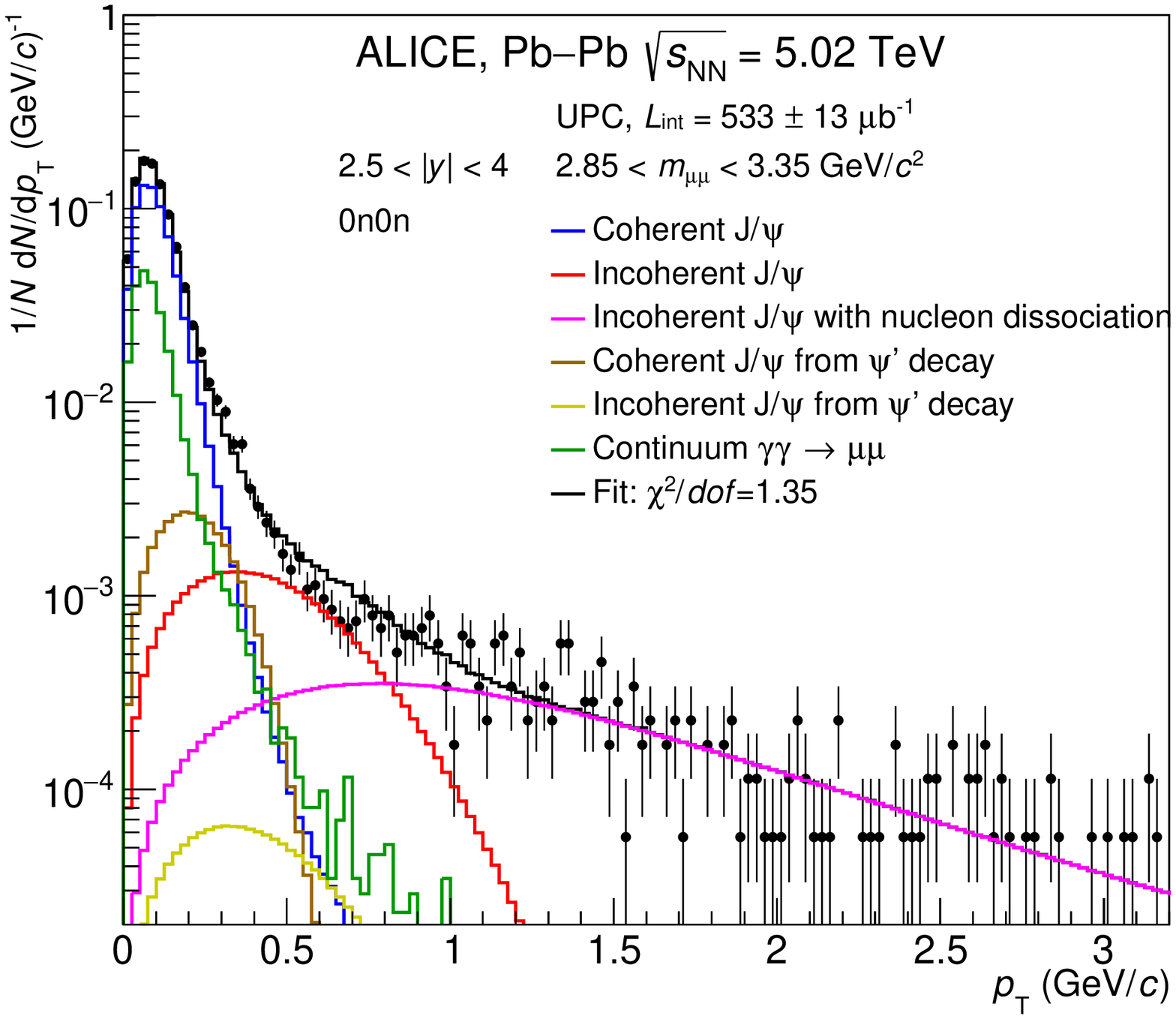}
\includegraphics[width=0.48\textwidth]{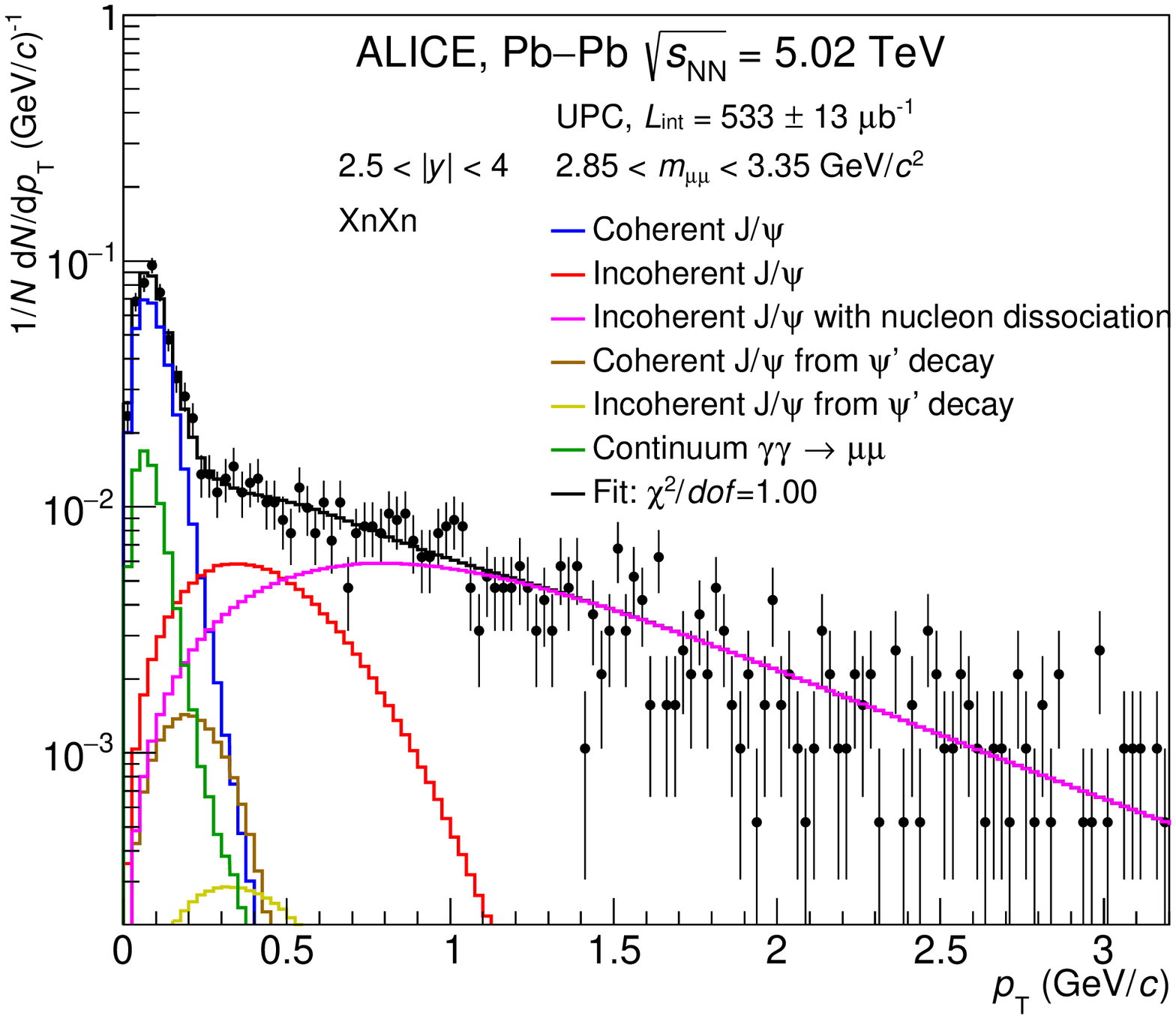}
\caption{ 
Transverse momentum distributions for events of the 0n0n (left) and XnXn (right) neutron classes measured at mid (top) and forward rapidity (bottom). The solid black markers represent data, the vertical line through each of them is the associated statistical uncertainty. The black lines depict the fit model described in the text.}
\label{fig:pt}
\end{figure}

The fraction $f_{\rm I}$, see Table~\ref{tab:upcXS}, is obtained from a fit to the transverse momentum distribution of lepton pairs in a restricted range of the invariant mass of the pair around the $\Jpsi$ pole mass ($2.9 < m < 3.2$ \GeVmass and $2.85 < m < 3.35$ \GeVmass for the analysis at mid and at forward rapidities, respectively).
This fit uses MC templates from STARlight for coherent and incoherent $\Jpsi$ production, $\Jpsi$ from decays of coherent and incoherent $\Psip$ production, and from the  $\ggll$ process. In addition, incoherent $\Jpsi$ production with nucleon dissociation, a process not included in the STARlight MC but present in data, is  taken into account, to describe the large transverse momentum region, with a template based on the H1 parameterisation of this process~\cite{H1:2013okq}. The normalisation of the templates from feed-down is constrained by the value of $f_{\rm D}$, while the normalisation of the $\ggll$ template is fixed by the fits to the invariant mass distribution. The three free parameters of the fit are the normalisation of both incoherent templates and of the coherent template. Examples of the fit to the transverse momentum  distributions for the mid and forward rapidity analyses for two different neutron classes are shown in Fig.~\ref{fig:pt}.

The last step to extract the $\Jpsi$ yield in a given class is to take into account the possibility of misclassification of events. There are two potential effects: that  a neutron is not measured (ZNA, ZNC efficiency) which could move an event, for example, from the Xn0n into the 0n0n class; and, that a neutron from a collision of an independent pair of Pb ions is detected (pile-up), which may then shift an event, for example, from the 0n0n to the Xn0n class.  The pile-up probability is determined using a special data sample collected with an unbiased trigger; it amounts to $0.0237\pm0.0005$ and $0.0238\pm0.0006$ for ZNA and ZNC, respectively. 
The efficiency of detecting at least one neutron in a Xn event is determined as explained in Ref.~\cite{ALICE:2022hbd} and amounts 
to $0.933 \pm 0.017$ for ZNA and $0.931 \pm 0.017$ for ZNC. The effects of pile-up and inefficiencies give rise to a migration matrix which is applied to the $N_{\rm coh}$ values of the different classes to obtain the $N_{\Jpsi}$ values. Note that migration does not change the total number of $\Jpsi$ candidates, so this correction introduces anti-correlated uncertainties. 

\subsection{Corrections
\label{sec:Cor}}

The term $ ({\rm A}\times\epsilon)$, in Eq.~(\ref{eq:xs}) is the product of three factors: $({\rm A}\times\epsilon)_{\det}$, $\epsilon_{\rm pu}$, and $\epsilon_{\rm emd}$. $({\rm A}\times\epsilon)_{\det}$ accounts for the acceptance and efficiency of the detector, including the trigger system, to measure the decay products of the coherently produced $\Jpsi$; it is obtained from the data sample of MC simulated events. 
These events were produced under the
assumption of transverse polarisation as expected for the case
of photoproduction and recently confirmed by the ALICE Collaboration~\cite{ALICE:2023svb}.
Polarisation could also play a role at midrapidity due to the interference of the two possible photon sources. As the incoming photon is linearly polarised, the interference causes an azimuthal anisotropy that can be observed in the final state~\cite{Xing:2020hwh, STAR:2022wfe}. Interference effects appear
at low values of transverse momenta and at midrapidity where both amplitudes are similar~\cite{Klein:1999gv}.  As the measurements presented here are integrated over transverse momentum, the potential impact of the interference is suppressed. Furthermore, the interference contributes mainly at small impact parameters, that is in the XnXn class, where our data sample has the largest statistical uncertainty and the contribution of interference effects is not visible.

The other two terms, $\epsilon_{\rm pu}$, and $\epsilon_{\rm emd}$, take into account the effects of pile-up. The first one, $\epsilon_{\rm pu}$, accounts for cases where, in addition to the coherent production of $\Jpsi$, another independent collision leaves signals in V0 or AD causing the event to be rejected at the trigger level. The pile-up probability is measured using data selected with an unbiased trigger based on the timing of bunches crossing the IP. For the midrapidity analysis $\epsilon_{\rm pu}=0.920\pm0.002$, while for the forward rapidity sample $\epsilon_{\rm pu}=0.962\pm0.001$; the uncertainty comes  from the size of the unbiased data sample. The second pile-up factor, $\epsilon_{\rm emd}$, takes into account events where the dissociation of the incoming nucleus produces, in addition to neutrons, charged particles that leave a signal in AD or V0. 
These extra particles come from EMD events with the neutron emission accompanied by the emission of protons or pions. According to Ref.~\cite{Pshenichnov:1999hw}, the corresponding cross sections are expected to be large. 
This factor is determined with a data sample triggered by an energy deposition over the threshold in either ZNA or ZNC; this sample is populated by EMD events~\cite{ALICE:2022hbd}.  For the 0n0n events $\epsilon_{\rm emd}=1.0$ as there is no nuclear dissociation. For the analysis at midrapidity $\epsilon_{\rm emd}=0.74\pm0.04$ and $\epsilon_{\rm emd}=0.57\pm0.05$ for the 0nXn+Xn0n and XnXn classes, respectively.    For the analysis at forward rapidity $\epsilon_{\rm emd}=0.88\pm0.01$ and $\epsilon_{\rm emd}=0.84\pm0.05$ for the Xn0n and XnXn classes, respectively. The uncertainty reflects the size of the data sample used to determine these factors.

\subsection{Systematic uncertainties}
A number of studies were undertaken to estimate potential systematic uncertainties. Their effect on the measured cross sections is summarised in
Tables~\ref{tab:uncCB} and~\ref{tab:uncMS}.

\begin{table}[t!]
\caption{
\label{tab:uncCB} 
Summary of the systematic uncertainties, given in percent, related to the  measurements performed with the central barrel detectors. The minus sign in the entry for migrations in the 0n0n class signifies that this uncertainty is anti-correlated with those from migrations in the 0nXn+Xn0n and XnXn classes. The second column identifies the type of uncertainty (U=uncorrelated, C=correlated, A=anticorrelated) as used in Eq.~(\ref{eq:Fit}).
 }
\centering
\begin{tabular}{lcrrrrrrr}
\hline
& & \multicolumn{3}{c}{$|y|<0.2$} & \multicolumn{3}{c}{$0.2<|y|<0.8$} \\
 Source & Type & 0n0n & 0nXn+Xn0n & XnXn& 0n0n & 0nXn+Xn0n & XnXn \\
\hline
 Signal extraction & U & 1.5 & 1.5 & 1.5& 1.5 & 1.5 & 1.5  \\ 
 Incoherent fraction & U&  0.1 & 1.5 & 1.3 & 0.1 & 1.5 & 1.3  \\ 
 Coherent shape & C& 0.1 & 0.8 & 0.6& 0.1 & 0.8 & 0.6  \\ 
  Feed-down & C& 0.6 & 0.6 & 0.6 & 0.6 & 0.6 & 0.6  \\ 	
 Branching ratio & C & 0.5 & 0.5 & 0.5 & 0.5 & 0.5 & 0.5  \\ 
   Luminosity & C & 2.5 & 2.5 & 2.5& 2.5 & 2.5 & 2.5  \\ 
 Trigger live time & C& 1.5 & 1.5 & 1.5& 1.5 & 1.5 & 1.5  \\ 
 ITS-TPC matching & C &2.8 & 2.8 & 2.8& 2.8 & 2.8 & 2.8  \\ 
 TOF trigger & C& 0.7 & 0.7 & 0.7& 0.7 & 0.7 & 0.7  \\ 
 SPD trigger & C&  1 & 1 & 1& 1 & 1 & 1 \\ 
 $\epsilon_{\rm pu}$ & C  & 3 & 3 & 3 & 3 & 3 & 3 \\ 
$\epsilon_{\rm emd}$  & C& 0 & 3.2 & 3.5 & 0 & 3.2 & 3.5  \\ 
 Migrations &A & $-3.9$  & 3.4 & 0.9& $-3.6$ & 3.1 & 1.1\\
\hline
\end{tabular}
\end{table}

\begin{table}[t!]
\caption{
\label{tab:uncMS} 
Summary of the systematic uncertainties, given in percent, related to the  measurements performed with the muon spectrometer. The minus sign in the entry for migrations in the 0n0n class signifies that this uncertainty is anti-correlated with those from migrations in the 0nXn+Xn0n and XnXn classes.
The second column identifies the type of uncertainty (U=uncorrelated, C=correlated, A=anticorrelated) as used in Eq.~(\ref{eq:Fit}).}
\centering
\begin{tabular}{lcrrrrrrrrr}
\hline
& & \multicolumn{3}{c}{$2.5<|y|<3.0$} & \multicolumn{3}{c}{$3.0<|y|<3.5$} & \multicolumn{3}{c}{$3.5<|y|<4.0$}\\
 Source & Type & 0n0n & Xn0n & XnXn& 0n0n & Xn0n & XnXn& 0n0n & Xn0n & XnXn \\
 \hline
 Signal extraction & U &0.2 &1.3 & 0.8 &0.1 & 0.6 & 0.7 & 0.5 & 0.5 & 0.9  \\ 
 Incoherent fraction & U & 0.4 & 0.6 & 1.6& 0.4 &0.9 &3.3 & 0.4 & 0.5 & 2.2  \\ 
 Coherent shape & C &0.1 & 0.1 & 0.1 & 0.1 & 0.1 & 0.1 & 0.1 & 0.1 & 0.1  \\ 
 Feed-down & C &0.7 & 0.7 & 0.7& 0.7 & 0.7 & 0.7& 0.7 & 0.7 & 0.7  \\ 
 Branching ratio & C & 0.6 & 0.6 & 0.6& 0.6 & 0.6 & 0.6& 0.6 & 0.6 & 0.6  \\ 
 Luminosity & C &2.5 & 2.5 & 2.5& 2.5 & 2.5 & 2.5& 2.5 & 2.5 & 2.5  \\ 
 Tracking & C & 3 & 3 & 3& 3 & 3 & 3 & 3 & 3 & 3  \\ 
 Trigger & C & 6.2 & 6.2 & 6.2& 6.2 & 6.2 & 6.2& 6.2 & 6.2 & 6.2  \\ 
 Matching & C& 1 & 1 & 1& 1 & 1 & 1& 1 & 1 & 1  \\ 
$\epsilon_{\rm pu}$ &C& 0.2 & 0.2 & 0.2& 0.2 & 0.2 & 0.2& 0.2 & 0.2 & 0.2  \\ 
 $\epsilon_{\rm emd}$ &C& 0 & 1.1 & 6 & 0 & 1.1 & 6 & 0 & 1.1 & 6  \\ 
Migrations & A& $-0.3$ & 3.8 & 3.3 &$-0.2$ & 3.6 & 3.6 & $-0.2$ & 3.3 & 3.6\\
\hline
\end{tabular}
\end{table}

To study the uncertainty on the model used for the signal extraction at midrapidity, the yield according to the Crystal Ball function is compared to counting the events under the peak region after the background is subtracted using the exponential shape from the fit.
The  model based on the Crystal Ball function is used as the baseline and  half of the difference, amounting to 1.5\%, is assigned as the systematic uncertainty.
 Another contribution to the uncertainty on the  signal extraction comes from the description of the background. This was estimated  by varying the fit range which produces a 0.3\% effect which is added in quadrature to the uncertainty on the modelling of the signal.
For the analysis at forward rapidity the uncertainty is estimated by varying the values of the tail parameters of the Crystal Ball function in the ranges found by fits to the signal in the simulated MC samples.
This uncertainty varies from 0.1\% to 1.3\% and is considered uncorrelated across rapidity and neutron classes.

There is an uncorrelated source of uncertainty for the determination of $f_{\rm I}$ that originates in the modelling of the different templates needed for the fit to the transverse momentum distribution described in Sec.~\ref{sec:YE}.
For the midrapidity analysis, it is estimated by using for the template of the $\ggll$ process either the transverse momentum distribution obtained at either side of the $\Jpsi$ peak  in the invariant mass distribution or the template from the STARlight MC. For the forward analysis, the shape of the incoherent distribution is obtained either from the fit described in Sec.~\ref{sec:YE} or it is constrained by fitting the transverse momentum distribution of the 0nXn sample with the requirement of activity in the ADC detector; this sample is dominated by incoherent production. The uncorrelated  uncertainty for  $f_{\rm I}$  varies from a fraction of a percent to a few percent.

There is also a correlated uncertainty related to the extraction of the incoherent contamination. It is known that the STARlight MC does not describe correctly the shape of the transverse momentum distribution for the coherent production of $\Jpsi$~\cite{ALICE:2021tyx}. A different shape for the transverse momentum dependence of coherent production was used in the fit and half the difference in the results is assigned as an uncertainty. The effect is below 1\% and it is larger for the midrapidity analysis since the resolution of the muon spectrometer is not as good as for the central barrel detectors, so it is not so sensitive to this effect.

The uncertainty on feed-down is estimated by varying $f_{\rm D}$ within its uncertainty. As the determination of feed-down is independently done using the central barrel detectors and the muon spectrometer,  this uncertainty is correlated only across the corresponding measurements and it is uncorrelated between the results obtained at mid and forward rapidities.  It amounts to 0.6\% and 0.7\%, respectively.

Two other sources of uncertainties are also considered as correlated: the uncertainty on the branching ratios is obtained from Ref.~\cite{ParticleDataGroup:2022pth}; the uncertainty on the determination of the luminosity, coming from the measurement of the reference cross sections
and from the stability of the calibration over time,
is taken from Ref.~\cite{ALICE:2022xir} and amounts to 2.5\%. For CBtrig there is another source of uncertainty related to the luminosity of the data sample, namely the precision to which the live time of the trigger is known, which is 1.5\%. The live time of MStrig is known with a very good precision and produces a negligible uncertainty.
 
 For the central barrel analysis there are four uncertainty sources that are correlated across the corresponding measurements.  A systematic uncertainty on the tracking efficiency of 2\% per track is estimated by comparing, in data and in MC,
 the matching efficiency for track segments reconstructed in the TPC and in the ITS. This leads to a 2.8\% systematic uncertainty for two tracks. The uncertainty of the TOF trigger efficiency due to the spread of the arrival times of various particle species to TOF is evaluated as 0.5\% per track (1\% in total). The uncertainty associated with the determination of the trigger efficiency of the SPD is obtained directly from data by varying the requirements to select the tracks used to measure this efficiency. This uncertainty amounts to 1\%.

There are three correlated uncertainties associated with the muon spectrometer.  The uncertainty on the tracking efficiency amounts to  3\%. It is estimated by comparing the single-muon tracking efficiency values obtained in MC and data, with a procedure that exploits the redundancy of the information from the tracking chambers~\cite{ALICE:2015jrl}. The systematic uncertainty on the dimuon trigger efficiency has two contributions.
 The uncertainty on the intrinsic efficiencies of the muon trigger chambers is determined by varying them in the MC by an amount equal to the statistical uncertainty on their measurement with a data-driven method and amounts to 1.5\%. The uncertainty on the response of the trigger algorithm is obtained by comparing the trigger response function between data and MC; it amounts to 6.0\%. These two contributions are added in quadrature. There is also a 1\% uncertainty on the matching efficiency of tracks reconstructed with the tracking and the trigger chambers.

The uncertainty  on $\epsilon_{\rm pu}$ for the forward rapidity analysis is obtained by varying this factor within its uncertainty. It amounts to 0.2\%.
As there are more detector systems contributing to $\epsilon_{\rm pu}$ in the CBtrig case, the uncertainty for the midrapidity measurements is estimated by repeating the analysis without the offline  veto from AD and V0, which increases both the yield and $\epsilon_{\rm pu}$. These increases do not compensate exactly and the ensuing difference of 3\% is assigned as a systematic uncertainty.
The uncertainty on $\epsilon_{\rm emd}$ is obtained by varying it within its uncertainty. It is of the order of a few percent, differing among the neutron classes and rapidity intervals.

The uncertainty on migrations across the different neutron classes is obtained by varying the pile-up in ZNA and ZNC as well as the efficiencies of these detectors within their uncertainties. At midrapidity the efficiency is the leading uncertainty for the XnXn neutron class, while pile-up dominates the other two neutron classes. At forward rapidities the efficiency is the leading uncertainty. The largest difference, with respect to the nominal measurement, from all  variations is taken as the uncertainty.  As mentioned above, these uncertainties are  anti-correlated across the neutron classes within one rapidity range.

\section{Results
\label{sec:Results}}

\subsection{Cross section in UPC
\label{sec:UPC}}
\begin{table}[t!]
\caption{
\label{tab:upcXS} 
Values for the number of $\Jpsi$ candidates ($N_{\rm fit}$), the incoherent fraction ($f_{\rm I}$), correction for the detector acceptance and efficiency ($({\rm A}\times\epsilon)_{\det}$) and the measured cross section (${\rm d}\sigma_{\rm PbPb}/{\rm d}y$) for the different neutron classes and rapidity ranges. The first uncertainty in the last column is statistical, the rest are systematic. The second uncertainty is uncorrelated, the third correlated, and the fourth originates from migrations across neutron classes. Note that for each rapidity range the 0n0n uncertainty related to migrations is preceded by a $\mp$, while the other neutron classes have a $\pm$; this means that these uncertainties are anti-correlated.
 }
\centering
\begin{tabular}{lrccc}
\hline
\multicolumn{1}{c}{Class} & \multicolumn{1}{c}{$N_{\rm fit}$}& $f_{\rm I}$ &  $({\rm A}\times\epsilon)_{\det}$ & ${\rm d}\sigma_{\rm PbPb}/{\rm d}y$ (mb) \\
\hline
 $|y|<0.2$ & & & &  \\
0n0n & $1744 \pm 49$ &$0.014\pm0.002$ & 0.053 &  $3.130 \pm 0.090  \pm 0.047  \pm 0.164 \mp 0.122$\\
0nXn+Xn0n & $412 \pm 23$ & $0.179\pm0.011$&0.053 & $0.730 \pm 0.050  \pm 0.015  \pm 0.045 \pm 0.025$\\
XnXn & $84 \pm 11$ &$0.144\pm0.021$ & 0.053& $0.250 \pm 0.024  \pm 0.005  \pm 0.016 \pm 0.002$ \\
\hline
 $0.2<|y|<0.8$ & & & &  \\
0n0n & $2179\pm 54$&$0.014\pm0.002$ & 0.024 & $2.900 \pm 0.070  \pm 0.044  \pm 0.152 \mp 0.104$\\
0nXn+Xn0n & $597\pm 28$ & $0.179\pm0.011$&0.024 &  $0.800 \pm 0.040  \pm 0.017  \pm 0.050 \pm 0.025$\\
XnXn & $134 \pm 13$ &$0.144\pm0.021$ & 0.024&  $0.300 \pm 0.029  \pm 0.006  \pm 0.019 \pm 0.003$\\
\hline
 $2.5<|y|<3.0$ & & & &  \\
0n0n & $2939 \pm 84$ & $0.0140 \pm 0.0038$ & 0.069 & $2.668 \pm 0.076  \pm 0.011  \pm 0.199 \mp 0.009$\\
Xn0n & $318 \pm 28$ & $0.070 \pm 0.0061$ & 0.069 &  $0.242 \pm 0.021  \pm 0.003  \pm 0.018 \pm 0.009$ \\
XnXn & $247 \pm 23$ & $0.1180 \pm 0.0159$& 0.069& $0.256 \pm 0.024  \pm 0.005  \pm 0.024 \pm 0.009$ \\
\hline
 $3.0<|y|<3.5$ & & & &  \\
0n0n & $7102 \pm 102$ & $0.0130 \pm 0.0041$& 0.194 & $2.322 \pm 0.033  \pm 0.010  \pm 0.173 \mp 0.005$\\
Xn0n &  $638 \pm 37$ & $0.0480 \pm 0.0090$ & 0.194 &  $0.172 \pm 0.010  \pm 0.002  \pm 0.013 \pm 0.006$\\
XnXn & $450 \pm 32$ & $0.1590 \pm 0.0332$& 0.194& $0.161 \pm 0.011  \pm 0.005  \pm 0.015 \pm 0.006$ \\
\hline
 $3.5<|y|<4.0$ & & & &  \\
0n0n & $2403 \pm 74$ &  $0.0070 \pm 0.0037$ & 0.097 &  $1.590 \pm 0.049  \pm 0.010  \pm 0.119 \mp 0.003$\\
Xn0n & $189 \pm 16$ &  $0.0270 \pm 0.0053$& 0.097 & $0.101 \pm 0.009  \pm 0.001  \pm 0.008 \pm 0.003$ \\
XnXn & $111 \pm 16$ &  $0.1650 \pm 0.0223$ & 0.097& $0.079 \pm 0.011  \pm 0.002  \pm 0.008 \pm 0.003$\\
\hline
\end{tabular}
\end{table}

Using the analysis strategy described in Sec.~\ref{sec:Ana}, the cross section for the coherent production of $\Jpsi$ vector mesons in UPCs of Pb nuclei at \fivenn is obtained. The measurements are reported in Table~\ref{tab:upcXS} along with other numerical values needed in 
Eq.~(\ref{eq:xs}) and~(\ref{eq:Ncoh}).

\begin{figure}[!t]
\centering
\includegraphics[width=0.48\textwidth]{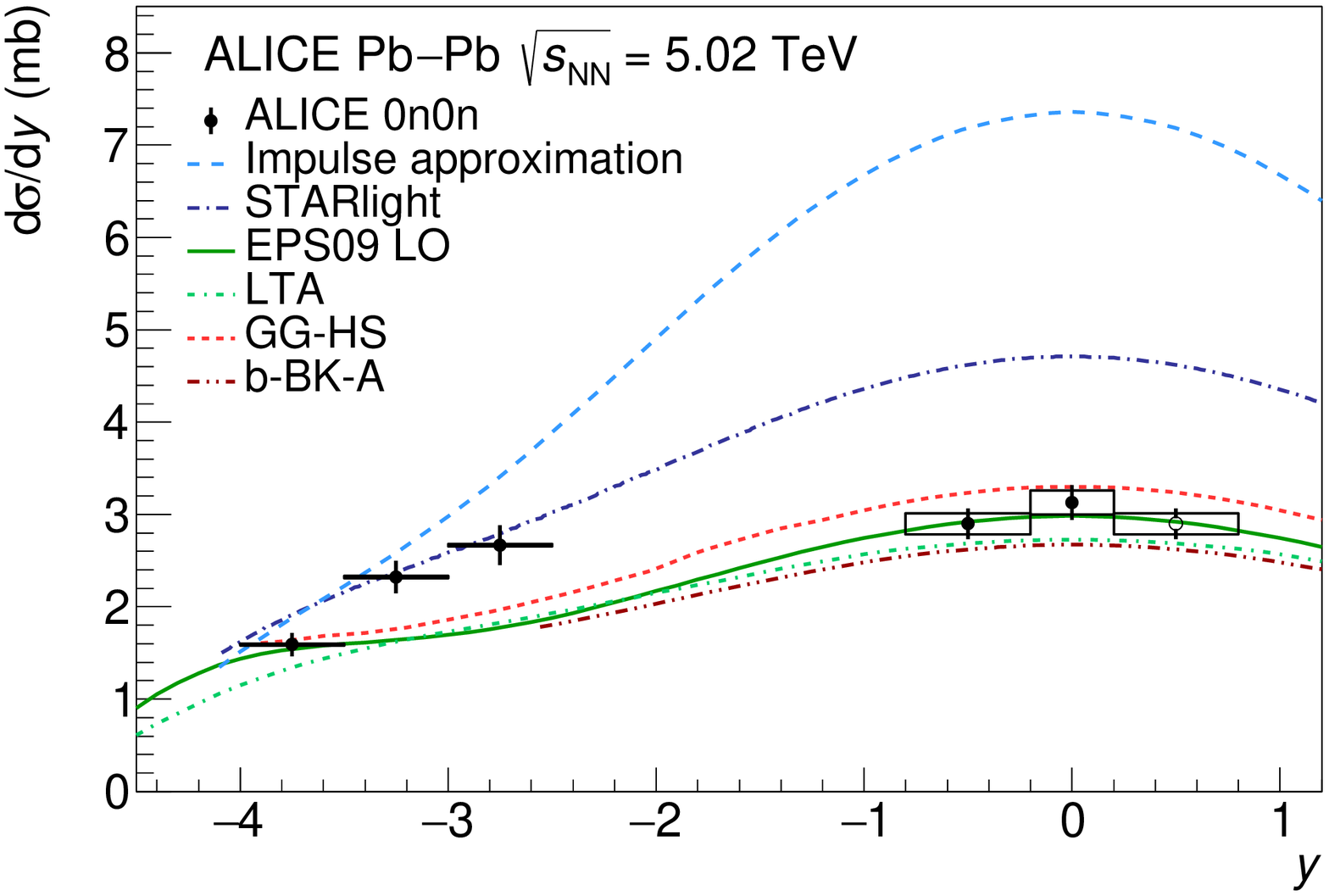}
\includegraphics[width=0.48\textwidth]{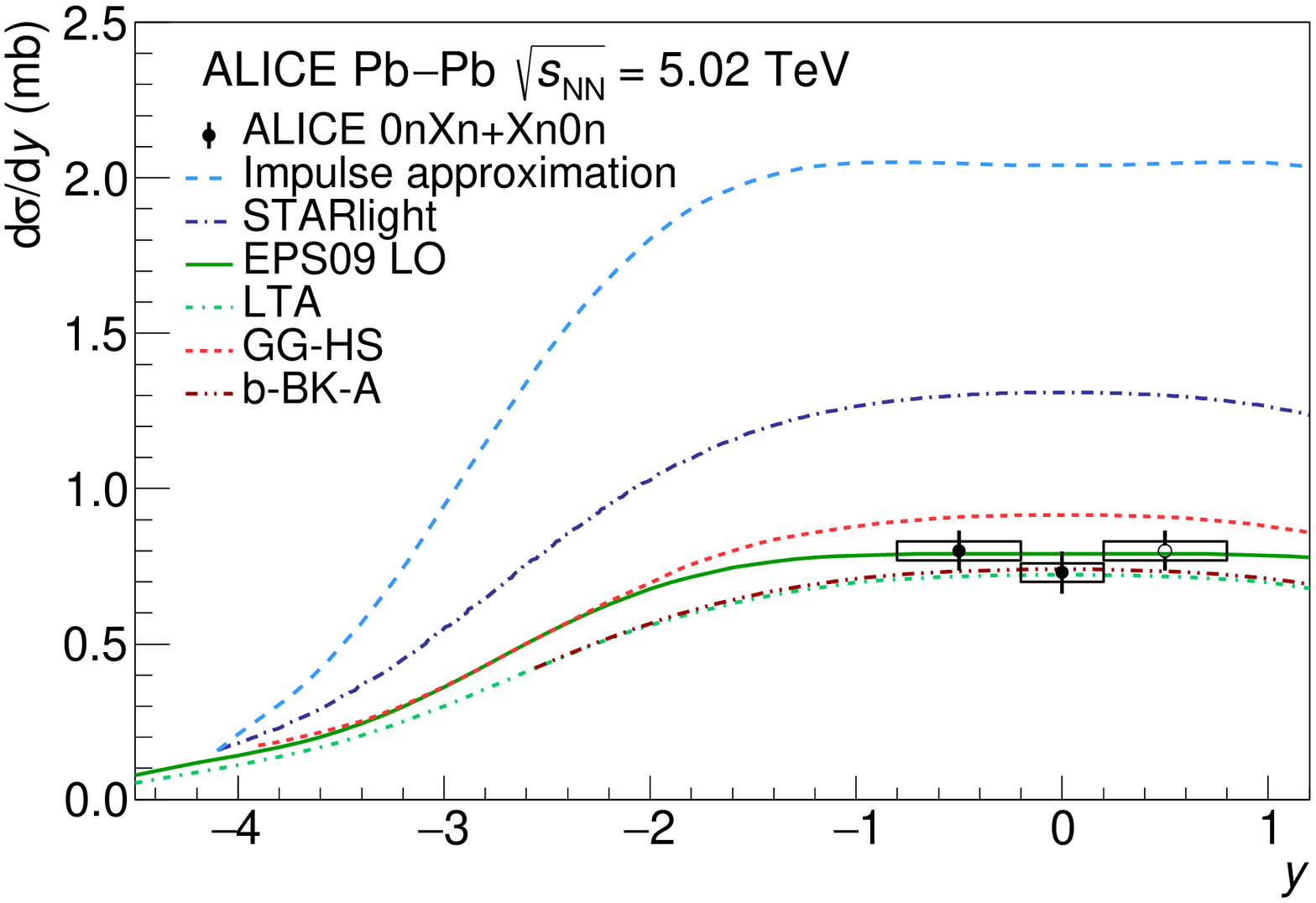}\\
\includegraphics[width=0.48\textwidth]{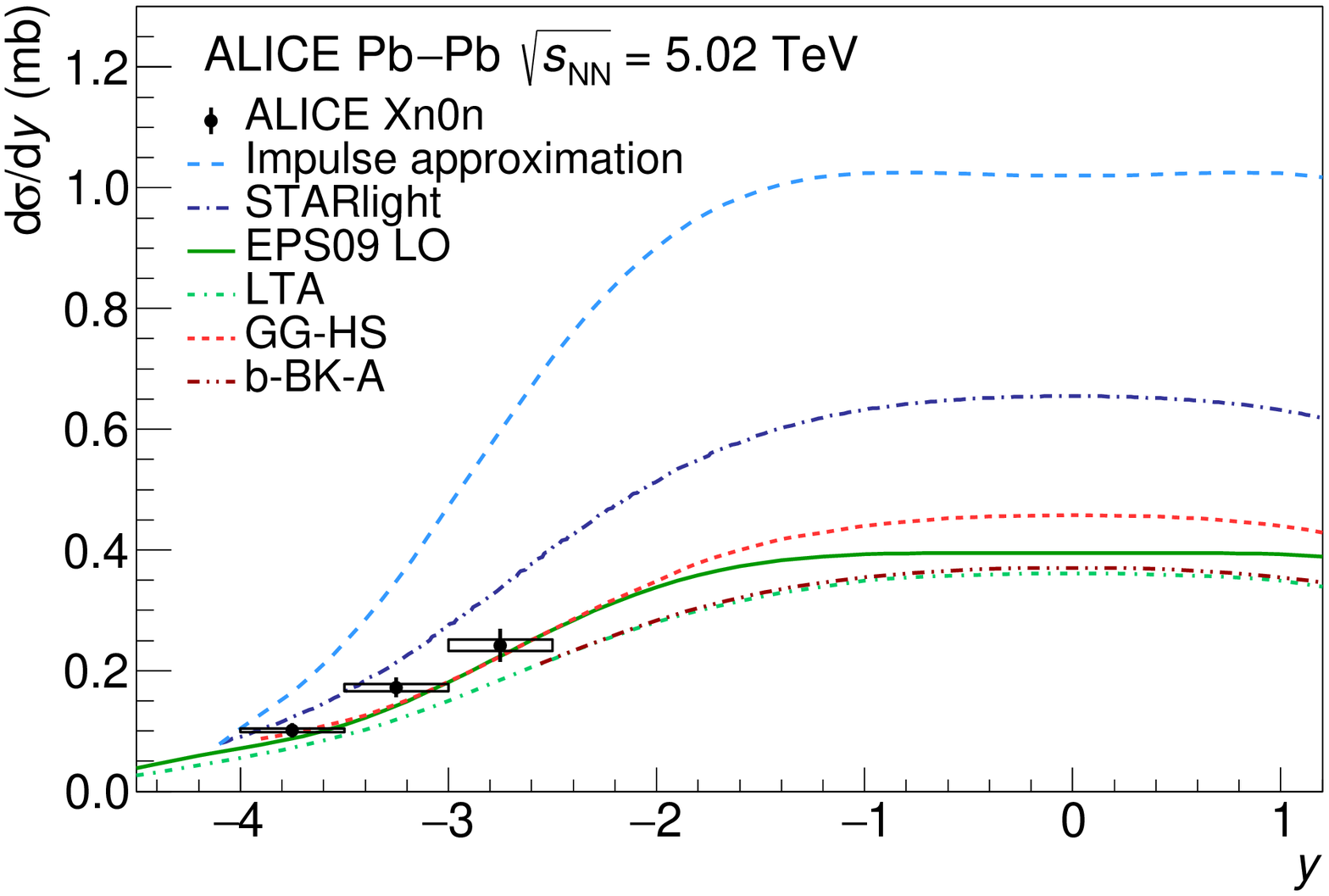}
\includegraphics[width=0.48\textwidth]{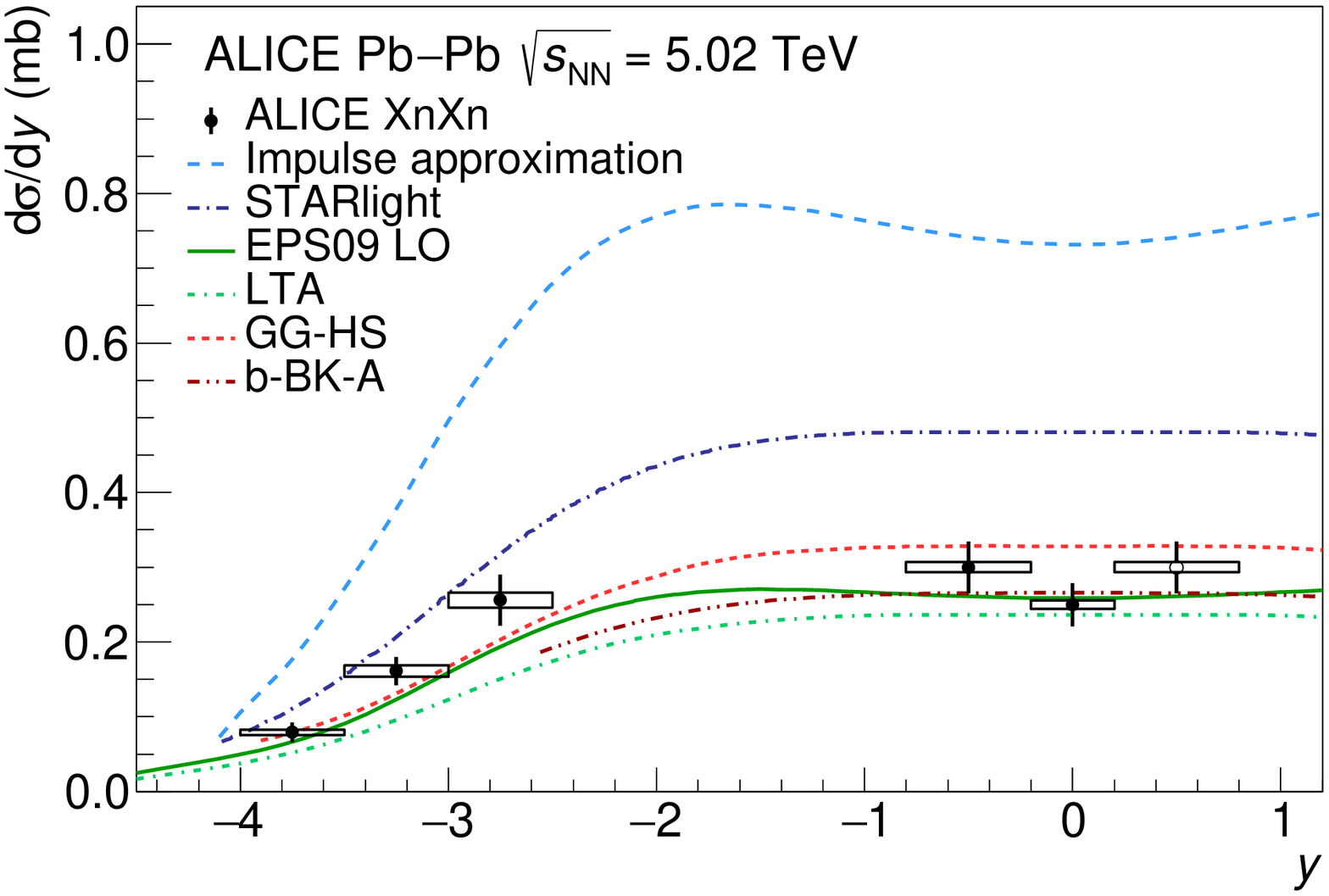}
\caption{
Measured cross section for the coherent  production of $\Jpsi$ in UPCs at \fivenn. The solid markers represent the measured cross section (the measurement at $0.2<\lvert y\rvert <0.8$ is shown at negative rapidities and reflected into positive rapidities with an open marker). The vertical line across a marker is the sum in quadrature of the statistical and uncorrelated systematic uncertainty. The width of the boxes depicts the range in rapidity covered by each measurement, while the height of a box is the sum in quadrature of the correlated systematic uncertainties and the effect of migrations     across neutron classes. Note that the uncertainties from migrations are anti-correlated between the 0n0n and the other two neutron classes in each rapidity interval. The lines depict the prediction of the different models discussed in Sec.~\ref{sec:Models}.}
\label{fig:upcXS}
\end{figure}

The results are compared to the predictions from different models in Fig.~\ref{fig:upcXS}.  The theoretical
uncertainties associated to each model are discussed in Sec.~\ref{sec:Models}. In all cases the IA calculation is well above the data signalling 
 important shadowing effects that have
 a   similar magnitude in  each  of the different neutron emission classes. The STARlight model describes well the data at forward rapidity in the 0n0n class, while overestimating the data in all  the other classes. At midrapidity, this model does not describe the data in any of the neutron classes.
The predictions of the other four models---EPS09-LO, LTA, b-BK-A, and GG-HS---are qualitatively similar, while quantitatively they differ with the maximum spread given by the GG-HS and b-BK-A models, whose predictions differ by up to about 20\%. These four models describe the data reasonably well, except for the rapidity range $3.5<|y|<2.5$ in the 0n0n class, where the data are clearly above the predictions. For the XnXn neutron class the data at forward rapidity are systematically slightly above the predictions.  

The IA and STARlight models do not include gluon shadowing or saturation effects. The EPS09-LO and LTA models do not include explicitly gluon saturation, while the b-BK-A and GG-HS predictions do not include explicitly shadowing effects beyond saturation. The data indicate that the IA and STARlight predictions are disfavoured, implying the need of some QCD dynamic effect, beyond what is included in these models, to describe the measurements. Within the experimental precision and the large theoretical uncertainties mentioned in Sec.~\ref{sec:Models},  models that include either shadowing or gluon saturation give an equally good description of the data.

\subsection{Extraction of the photonuclear cross section
\label{sec:gPb}}

Having several independent UPC measurements allows for the extraction of the two photonuclear cross sections in each rapidity interval.
A $\chi^2$ minimisation is applied to the three measurements in each $y$ range.  The used $\chi^2$ approach incorporates  the correlated uncertainties through nuisance parameters and the uncorrelated and statistical ones utilising relative uncertainties. This method has already been used by the ALICE Collaboration to extract the energy dependence of  exclusive photoproduction of $\Jpsi$ in p--Pb collisions~\cite{ALICE:2018oyo,ALICE:2023mfc} and it was originally used by the H1 Collaboration for the measurement of the inclusive deeply inelastic cross section at HERA~\cite{H1:2009jxj}. The $\chi^2$ definition is given by

\begin{equation}
 \chi^2_{\rm exp}\left(\vec{m},\vec{b}\right) = 
 \sum_i
 \frac{\left(m^i
- \sum_j \gamma^{\, i}_j m^i b_j  - {\mu^i} \right)^2}
{ \textstyle \delta^2_{i,{\rm stat}}\mu^i\left(m^i -  \sum_j \gamma^{\, i}_j m^i b_j\right)+
\left(\delta_{i,{\rm uncor}}\,  m^i\right)^2}
 + \sum_j b^2_j.
\label{eq:Fit}
\end{equation}
Here, ${\mu^i}$ is the measured central value  at a point $i$,
 $m^i$ are  given by the right-hand side of Eq.~(\ref{eq:upc}) with the fluxes computed with the \Noon\ program~\cite{Broz:2019kpl} (see Sec.~\ref{sec:Flux} and Table~\ref{tab:flux} for the flux values), and $\sigma_{\gPb}(\pm y)$  the two parameters to be extracted from the fit. Note that for the rapidity range $|y|<0.2$ there is only one photonuclear cross section. The relative statistical and uncorrelated systematic uncertainties for each rapidity range (see Tables~\ref{tab:uncCB} and~\ref{tab:uncMS}) are given by $\delta_{i,{\rm stat}} = \Delta_{i, {\rm stat}}/\mu^i$ and 
$\delta_{i,{\rm uncor}} = \Delta_{i,{\rm uncor}}/\mu^i$,
respectively. Finally, $\gamma^i_j$ is the matrix of 
 the correlated systematic uncertainty for  the source of type $j$ at the point $i$, where $b_j$ is the  associated set of nuisance parameters.

\begin{table}[t!]
\caption{
\label{tab:flux} 
 Theoretical input needed to obtain the photonuclear cross section and the nuclear suppression factor. Photon fluxes, see Eq.~(\ref{eq:upc}), computed with \Noon \ for the different neutron classes and rapidity ranges. The last column shows the value of $\sigma^{\rm IA}_{\gPb}$ as computed in Ref.~\cite{Guzey:2013xba}. 
 }
 \centering
\begin{tabular}{rrrrr}
\hline
 $y$ & $n_\gamma$(0n0n) &  $n_\gamma$(0nXn+Xn0n) &  $n_\gamma$(XnXn) & $\sigma^{\rm IA}_{\gPb}$ ($\mu$b)\\
\hline
 $3.5<y<4$& 178.51  & 18.18  & 6.34 &10\\ 
$3<y<3.5$ & 162.99  & 18.19  & 6.34  & 14\\ 
 $2.5<y<3$ & 147.46  & 18.19  & 6.34   & 19\\ 
 $0.2<y<0.8$ & 77.88  & 17.88  & 6.33   & 48\\ 
 $-0.2<y<0.2$ & 62.86  & 17.47  & 6.27   & 58\\ 
$-0.8<y<-0.2$ & 48.31  & 16.75    & 6.18 & 71\\ 
$-3<y<-2.5$ & 3.91  & 4.97  & 2.78  & 176\\ 
$-3.5<y<-3$  & 1.22  & 2.15  & 1.42  & 215\\ 
$-4<y<-3.5$  & 0.26  & 0.61  & 0.48  & 262\\ 
\hline
\end{tabular}
\end{table}

\begin{table}[t!]
\caption{
\label{tab:gPb} 
Photonuclear cross sections extracted from the UPC measurements using the procedure described in the text. The quoted  uncertainties are uncorrelated (unc.), correlated (corr.), caused by migrations across neutron classes (mig.) and by variations of the flux fractions in the different classes (flux frac.). The lines separate the different ranges in $|y|$.
Note that  two photonuclear cross sections in each rapidity interval are anti-correlated. 
 }
 \centering
\begin{tabular}{crrrrrr}
\hline
$y$& $\WgPb$ (\GeV) & $\sigma_{\gPb}$ ($\mu$b) & unc. ($\mu$b)  & corr. ($\mu$b) & mig. ($\mu$b) & flux frac. ($\mu$b)   \\
\hline
$3.5<y<4$& 19.12 &8.84 &0.30 &0.68 &0.02 &0.04  \\
$-4<y<-3.5$& 813.05 &57.32 &20.77 &7.57 &6.41 &6.56  \\
\hline
$3<y<3.5$ & 24.55 &13.89 &0.23 &1.08 &0.05 &0.08  \\
$-3.5<y<-3$ & 633.21 &46.58 &6.61 &5.73 &3.77 &3.63  \\
\hline
$2.5<y<3$ & 31.53 &16.89 &0.59 &1.32 &0.11 &0.18  \\
$-3<y<-2.5$ & 493.14 &44.68 &6.38 &5.15 &2.73 &2.97  \\
\hline
 $0.2<y<0.8$ & 97.11 &21.73 &5.12 &3.12 &4.32 &2.73  \\
$-0.8<y<-0.2$ & 160.10 &25.00 &7.33 &4.88 &5.43 &3.91  \\
\hline
$-0.2<y<0.2$ & 124.69 &24.15 &0.69 &1.37 &0.50 &0.06  \\
\hline
\end{tabular}
\end{table}

\begin{figure}[!t]
\centering
\includegraphics[width=0.98\textwidth]{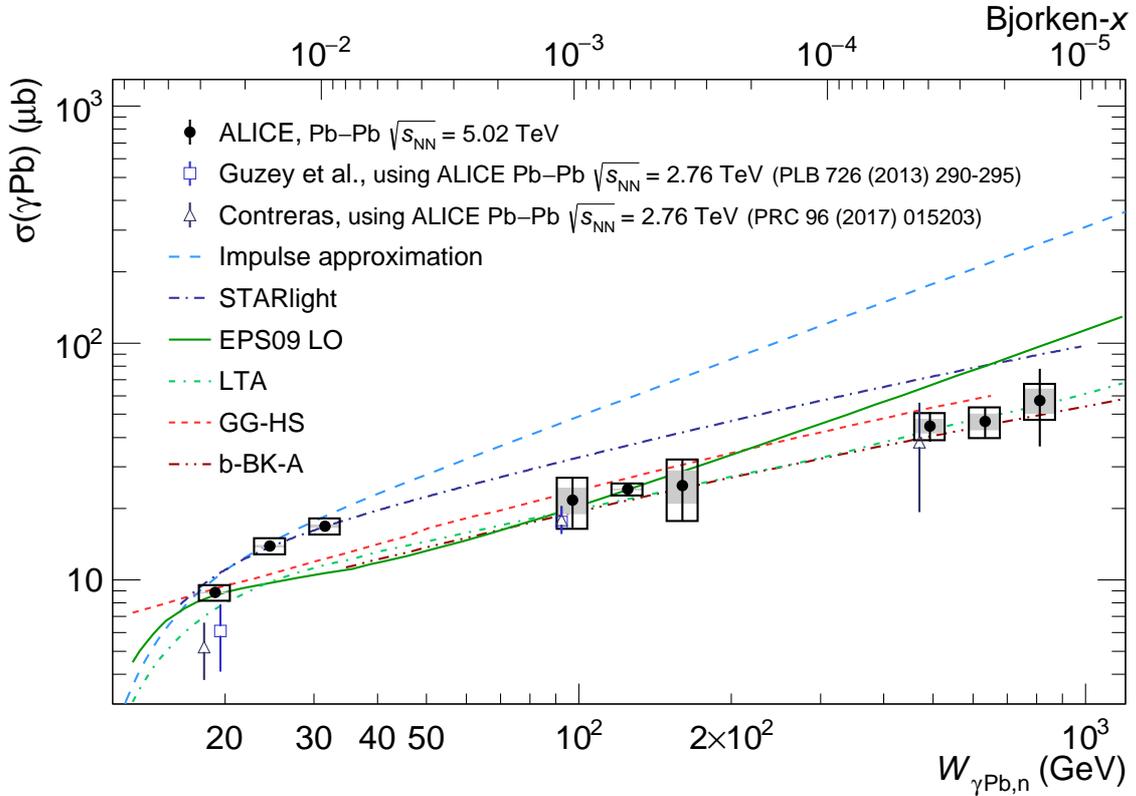}
\caption{Photonuclear cross section for the $\gamma+{\rm Pb}\to \Jpsi+{\rm Pb}$ process as a function of $\WgPb$ (lower axis) or Bjorken-$x$ (upper axis). The solid markers represent the measured cross section. The vertical line across a marker is the uncorrelated uncertainty. The height of an empty box is the sum in quadrature of the correlated systematic uncertainties and the effect of migrations across neutron classes. The gray box represents the theoretical uncertainty coming from the computation of the photon flux.  The lines depict the prediction of the different models discussed in Sec.~\ref{sec:Models}. The open  triangular and square markers show the cross sections extracted in Refs.~\cite{Contreras:2016pkc,Guzey:2013xba} using ALICE Run~1 data.}
\label{fig:gPb}
\end{figure}

The uncertainties for the measurement of $\sigma_{\gPb}(y)$ are obtained as follows. A fit including the statistical as well as the correlated and uncorrelated systematic uncertainties is performed. Another fit including only the statistical and uncorrelated systematic uncertainties is performed. The uncertainty from this second fit is quoted as the uncorrelated uncertainty. 
The difference between the uncertainties from the first and second fit, taken in quadrature, are quoted. 

There are two contributions to the uncertainty associated to the photon fluxes. One is related to the total flux and the other to 
the fractions of the total flux in each neutron class. The first contribution  is obtained by varying
the parameter of the nuclear radius in the  Woods--Saxon distribution according to neutron-skin measurements~\cite{Abrahamyan:2012gp}; this uncertainty amounts to
2\%  correlated over all rapidity intervals and neutron classes. This factor is already taken into account in the correlated uncertainties mentioned in the previous paragraph.  The second contribution is estimated by varying by $\pm 5$\% all cross sections used as input in \Noon\ for the computation of the photon flux fractions (see also Ref.~\cite{Baltz:1996as}). The relative change in the photon fluxes goes from 1\% to 8\% depending on rapidity and neutron class. These changes are anti-correlated in neutron classes for each rapidity interval. To compute the associated uncertainties, fits---including the statistical, correlated and uncorrelated systematic uncertainties---are performed using the modified fluxes. The largest difference, divided by $\sqrt{2}$, between these fits and the  fit with the default photon-flux values from \Noon\ is taken as the uncertainty originating from the photon flux. 
If the fluxes of STARlight were used, instead of those from \Noon, then
the results would vary by less than one percent, except for the
two largest energies, where the cross sections would be larger by 2.6\%
and 7.7\% at $\WgPb=633$ GeV and  $\WgPb=813$ GeV, respectively. This is well
within the uncorrelated uncertainties of the measurement. 

Uncertainties caused by the migrations across neutron classes are treated in a similar way to those associated with the photon flux. The input UPC cross sections are modified by the migration uncertainties, new fits are performed and the largest difference, divided by $\sqrt{2}$, with respect to the fit that uses the unmodified UPC cross sections is taken as the systematic uncertainty due to migration effects.

The results obtained by following this procedure are listed in Table~\ref{tab:gPb} and shown in Fig.~\ref{fig:gPb}, where they are compared to the predictions of the different models. Note that according to 
Eq.~(\ref{eq:upc}) the results for the cross section at low and high $\WgPb$ in one rapidity interval are anti-correlated. Note that the uncertainties for the high $\WgPb$ region are large, reaching about 30\% at $\WgPb = 813$ \GeV. 
The predictions obtained with IA~\cite{Guzey:2013xba} are consistent with the data for the  energy region below 40 \GeV, although systematically above the data; at all other energies the predictions from IA are well above the measurements with the difference increasing with energy. STARlight predictions describe the data for energies below 40 \GeV, but overestimate the measurements at all other energies. 
None of the EPS09-LO, LTA, b-BK-A, and GG-HS models describe the data in the  $\WgPb$ range from about 25 to 35 \GeV.  The EPS09-LO model describes the measurements at the lowest energy and at intermediate energies, but overestimates the measurements at the highest energies.  The GG-HS model does not include the reduction of phase space at low $\WgPb$, but  it  describes the data, except for the mentioned energy range, for all other measurements, with the predictions systematically on the higher side of the measurements. The predictions of the LTA and b-BK-A models are very similar and describe the data fairly well at all energies,  except for the energy range from about 25 to 35 \GeV.

 The photonuclear cross sections extracted in Refs.~\cite{Contreras:2016pkc,Guzey:2013xba} using ALICE Run~1 data are also shown in Fig.~\ref{fig:gPb}. The cross sections at the two highest $\WgPb$, namely 92 \GeV and 470 \GeV, agree with the new measurements presented here, while the two cross sections at low $\WgPb$ are below the new measurements by around 1.5 standard deviations. The fact that the cross sections extracted using the peripheral and ultra-peripheral results from Run 1 and the new measurements presented here agree reasonably well is remarkable, because they involve a different set of systematic uncertainties. It is also worth noting that the new measurements extend the range in $\WgPb$ by about 350 \GeV, up to $\WgPb=813$ \GeV, with respect to the maximum energy reached by ALICE Run~1 data.

As mentioned above, the CMS Collaboration submitted results on this
process~\cite{CMS:2023snh}. The CMS data
cover the ranges around 40 GeV to 50 GeV and 300 to 400 GeV in
$\WgPb$. These ranges lie in between the ranges covered by the ALICE forward and
midrapidity analyses. 
The results of the CMS Collaboration smoothly follow the same trend as the cross sections measured by ALICE. At low energies the measurements are compatible with the STARlight predictions and at high energies with the LTA and b-BK-A predictions.

\subsection{Nuclear suppression factor
\label{sec:nsp}}

\begin{figure}[!t]
\centering
\includegraphics[width=0.98\textwidth]{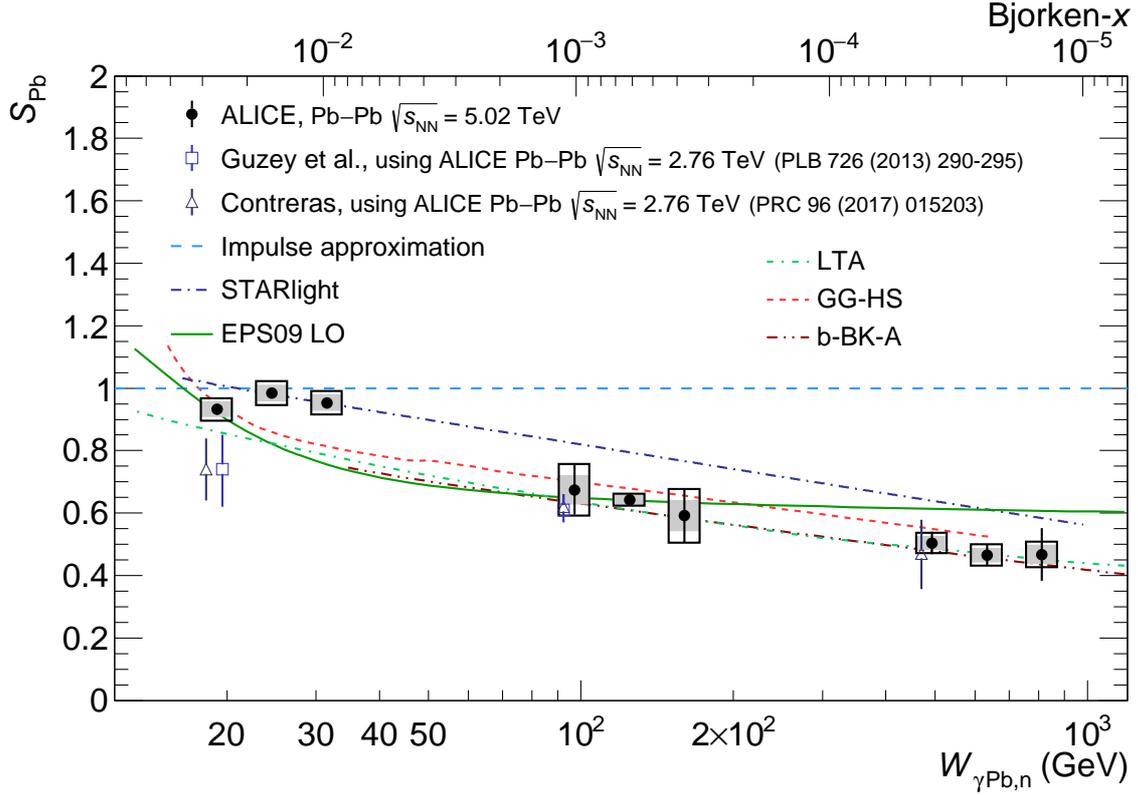}
\caption{
Nuclear suppression factor for the $\gamma+{\rm Pb}\to \Jpsi+{\rm Pb}$ process as a function of $\WgPb$ (lower axis) or Bjorken-$x$ (upper axis). The solid markers represent the measurement. The vertical line across a marker is the  uncorrelated uncertainty. The height of an empty box is the sum in quadrature of the correlated systematic uncertainties and the effect of migrations across neutron classes. A gray box represents the theoretical uncertainty coming from the computation of the photon flux and of the impulse approximation. The lines depict the prediction of the different models discussed in  Sec.~\ref{sec:Models}.
The open  triangular and square markers show the nuclear suppression factor extracted in Refs.~\cite{Contreras:2016pkc,Guzey:2013xba} using ALICE Run~1 data.}
\label{fig:nsf}
\end{figure}

The nuclear suppression factor is defined in Eq.~(\ref{eq:nsf}). 
To obtain it, the measured photonuclear cross sections are divided by the IA values, where we use the implementation from Ref.~\cite{Guzey:2013xba}. The corresponding values of IA are listed in Table~\ref{tab:flux}. 
According to Ref.~\cite{Guzey:2013xba} the computation of IA has an uncertainty of about 5\%, which reflects the uncertainties related to the experimental input data and its parameterisation. This uncertainty is taken into account in the results shown below.

The nuclear suppression factor is important because it provides a quantitative measure of shadowing in this process and several theoretical uncertainties, e.g. that associated to the $\Jpsi$ wave function, should largely cancel in the ratio.  Not all uncertainties cancel out completely; for example, in Ref.~\cite{Guzey:2013xba} it is argued that the interpretation of the nuclear suppression factor in terms of the gluon shadowing factor has a theoretical uncertainty due to corrections, amounting to about 10\%, that account for the  skewedness  and  the real part of the amplitude.

The nuclear suppression factor is shown in Fig.~\ref{fig:nsf}, where the measurement is compared with the predictions of the different models. The nuclear suppression factor at low energies is about 0.94, decreases to values slightly above 0.64 at intermediate energies, and decreases further down to about 0.47  at the highest measured energies. The  STARlight model describes only the $\WgPb$ range from about 25 to 35 \GeV. The other four models do not describe this energy range, but provide a fair description at higher energies, except for the EPS09-LO model, which predicts a  nuclear suppression factor that remains constant with increasing $\WgPb$, while the data and the other models exhibit a decreasing trend. The predictions of LTA and b-BK-A are quite close to each other and 
follow the behaviour of data at all energies, except for the  range from about 25 to 35 \GeV.

\section{Summary and outlook}

The coherent photonuclear production of $\Jpsi$ accompanied by electromagnetic  dissociation (EMD) was measured in a wide kinematic domain. Cross sections are reported for five rapidity intervals and three EMD classes. These measurements are used to extract the photonuclear cross section $\sigma_{\gPb}$ in the kinematic range 
$17 < \WgPb <920$ \GeV, which  corresponds to a Bjorken-$x$ interval of about three orders of magnitude: $ 1.1\times10^{-5}<x<3.3\times 10^{-2}$. In addition, the nuclear suppression factor was measured in the same energy range. These results, together with previous ALICE measurements, provide unprecedented information to probe quantum chromodynamics at high energies.

The results are compared to cross sections and nuclear suppression factors obtained from ALICE Run~1 data. The new measurements agree with the lower energy results, provide a large improvement in precision, and extend the reach of ALICE data in $\WgPb$, by about 350 \GeV, up to   $\WgPb=813$ \GeV.  

The results are also compared with different theoretical models. At low energies, predictions from IA are consistent with data, albeit systematically above them,  while with increasing energies IA overestimates data more and more, signalling important energy-dependent shadowing effects. The STARlight model describes the low energy data, but overestimates the measurements at large energies. As neither IA nor STARlight include shadowing or saturation effects, their comparison with data points to the presence of nuclear QCD phenomena at high energies beyond what is included in these two predictions. 
All other models considered---based on EPS09-LO, the leading-twist approximation, solutions of the impact-parameter dependent BK equation, and an energy-dependent hot-spot approach---describe correctly 
the data at high energy, correspondingly small Bjorken-$x$,  but underpredict the data in the $\WgPb$ range from about 25 to 35 \GeV. Within the uncertainties,  both saturation or shadowing models  give a reasonable description of data. The nuclear suppression factor varies from about 0.95 to 0.47 as the energy (Bjorken-$x$) increases (decreases). 

New data to be collected during Run 3 and 4 at the LHC will provide substantially larger data samples~\cite{Citron:2018lsq}, and the ALICE detector has been upgraded to fully exploit the new data sets~\cite{ALICE:2023udb}. These improvements should allow for more detailed studies and a reduction of the size of uncertainties. Furthermore, the measurement of the coherent photoproduction of $\Jpsi$ in peripheral \PbPb collisions will provide an alternative set of measurements, with different uncertainty sources, to extract the energy dependence of this process. Under these circumstances,  a global analysis of all data promises a measurement of the photonuclear cross section $\sigma_{\gPb}$ over three orders of magnitude in Bjorken-$x$ with a small experimental uncertainty.


\newenvironment{acknowledgement}{\relax}{\relax}
\begin{acknowledgement}
\section*{Acknowledgements}

The ALICE Collaboration would like to thank all its engineers and technicians for their invaluable contributions to the construction of the experiment and the CERN accelerator teams for the outstanding performance of the LHC complex.
The ALICE Collaboration gratefully acknowledges the resources and support provided by all Grid centres and the Worldwide LHC Computing Grid (WLCG) collaboration.
The ALICE Collaboration acknowledges the following funding agencies for their support in building and running the ALICE detector:
A. I. Alikhanyan National Science Laboratory (Yerevan Physics Institute) Foundation (ANSL), State Committee of Science and World Federation of Scientists (WFS), Armenia;
Austrian Academy of Sciences, Austrian Science Fund (FWF): [M 2467-N36] and Nationalstiftung f\"{u}r Forschung, Technologie und Entwicklung, Austria;
Ministry of Communications and High Technologies, National Nuclear Research Center, Azerbaijan;
Conselho Nacional de Desenvolvimento Cient\'{\i}fico e Tecnol\'{o}gico (CNPq), Financiadora de Estudos e Projetos (Finep), Funda\c{c}\~{a}o de Amparo \`{a} Pesquisa do Estado de S\~{a}o Paulo (FAPESP) and Universidade Federal do Rio Grande do Sul (UFRGS), Brazil;
Bulgarian Ministry of Education and Science, within the National Roadmap for Research Infrastructures 2020-2027 (object CERN), Bulgaria;
Ministry of Education of China (MOEC) , Ministry of Science \& Technology of China (MSTC) and National Natural Science Foundation of China (NSFC), China;
Ministry of Science and Education and Croatian Science Foundation, Croatia;
Centro de Aplicaciones Tecnol\'{o}gicas y Desarrollo Nuclear (CEADEN), Cubaenerg\'{\i}a, Cuba;
Ministry of Education, Youth and Sports of the Czech Republic, Czech Republic;
The Danish Council for Independent Research | Natural Sciences, the VILLUM FONDEN and Danish National Research Foundation (DNRF), Denmark;
Helsinki Institute of Physics (HIP), Finland;
Commissariat \`{a} l'Energie Atomique (CEA) and Institut National de Physique Nucl\'{e}aire et de Physique des Particules (IN2P3) and Centre National de la Recherche Scientifique (CNRS), France;
Bundesministerium f\"{u}r Bildung und Forschung (BMBF) and GSI Helmholtzzentrum f\"{u}r Schwerionenforschung GmbH, Germany;
General Secretariat for Research and Technology, Ministry of Education, Research and Religions, Greece;
National Research, Development and Innovation Office, Hungary;
Department of Atomic Energy Government of India (DAE), Department of Science and Technology, Government of India (DST), University Grants Commission, Government of India (UGC) and Council of Scientific and Industrial Research (CSIR), India;
National Research and Innovation Agency - BRIN, Indonesia;
Istituto Nazionale di Fisica Nucleare (INFN), Italy;
Japanese Ministry of Education, Culture, Sports, Science and Technology (MEXT) and Japan Society for the Promotion of Science (JSPS) KAKENHI, Japan;
Consejo Nacional de Ciencia (CONACYT) y Tecnolog\'{i}a, through Fondo de Cooperaci\'{o}n Internacional en Ciencia y Tecnolog\'{i}a (FONCICYT) and Direcci\'{o}n General de Asuntos del Personal Academico (DGAPA), Mexico;
Nederlandse Organisatie voor Wetenschappelijk Onderzoek (NWO), Netherlands;
The Research Council of Norway, Norway;
Commission on Science and Technology for Sustainable Development in the South (COMSATS), Pakistan;
Pontificia Universidad Cat\'{o}lica del Per\'{u}, Peru;
Ministry of Education and Science, National Science Centre and WUT ID-UB, Poland;
Korea Institute of Science and Technology Information and National Research Foundation of Korea (NRF), Republic of Korea;
Ministry of Education and Scientific Research, Institute of Atomic Physics, Ministry of Research and Innovation and Institute of Atomic Physics and University Politehnica of Bucharest, Romania;
Ministry of Education, Science, Research and Sport of the Slovak Republic, Slovakia;
National Research Foundation of South Africa, South Africa;
Swedish Research Council (VR) and Knut \& Alice Wallenberg Foundation (KAW), Sweden;
European Organization for Nuclear Research, Switzerland;
Suranaree University of Technology (SUT), National Science and Technology Development Agency (NSTDA), Thailand Science Research and Innovation (TSRI) and National Science, Research and Innovation Fund (NSRF), Thailand;
Turkish Energy, Nuclear and Mineral Research Agency (TENMAK), Turkey;
National Academy of  Sciences of Ukraine, Ukraine;
Science and Technology Facilities Council (STFC), United Kingdom;
National Science Foundation of the United States of America (NSF) and United States Department of Energy, Office of Nuclear Physics (DOE NP), United States of America.
In addition, individual groups or members have received support from:
European Research Council, Strong 2020 - Horizon 2020 (grant nos. 950692, 824093), European Union;
Academy of Finland (Center of Excellence in Quark Matter) (grant nos. 346327, 346328), Finland.

\end{acknowledgement}

\bibliographystyle{utphys}   
\bibliography{bibliography}

\newpage
\appendix

%
%

\section{The ALICE Collaboration}
\label{app:collab}
\begin{flushleft} 
\small

S.~Acharya\,\orcidlink{0000-0002-9213-5329}\,$^{\rm 126}$, 
D.~Adamov\'{a}\,\orcidlink{0000-0002-0504-7428}\,$^{\rm 87}$, 
A.~Adler$^{\rm 70}$, 
G.~Aglieri Rinella\,\orcidlink{0000-0002-9611-3696}\,$^{\rm 33}$, 
M.~Agnello\,\orcidlink{0000-0002-0760-5075}\,$^{\rm 30}$, 
N.~Agrawal\,\orcidlink{0000-0003-0348-9836}\,$^{\rm 51}$, 
Z.~Ahammed\,\orcidlink{0000-0001-5241-7412}\,$^{\rm 133}$, 
S.~Ahmad\,\orcidlink{0000-0003-0497-5705}\,$^{\rm 16}$, 
S.U.~Ahn\,\orcidlink{0000-0001-8847-489X}\,$^{\rm 71}$, 
I.~Ahuja\,\orcidlink{0000-0002-4417-1392}\,$^{\rm 38}$
% ...

\section*{Affiliation Notes}

$^{\rm I}$ Also at: Max-Planck-Institut f\"{u}r Physik, Munich, Germany\\
$^{\rm II}$ Also at: Italian National Agency for New Technologies, Energy and Sustainable Economic Development (ENEA), Bologna, Italy\\
$^{\rm III}$ Also at: Department of Applied Physics, Aligarh Muslim University, Aligarh, India\\
$^{\rm IV}$ Also at: Institute of Theoretical Physics, University of Wroclaw, Poland\\
$^{\rm V}$ Also at: An institution covered by a cooperation agreement with CERN\\

\section*{Collaboration Institutes}

$^{1}$ A.I. Alikhanyan National Science Laboratory (Yerevan Physics Institute) Foundation, Yerevan, Armenia\\
$^{2}$ AGH University of Science and Technology, Cracow, Poland\\
$^{3}$ Bogolyubov Institute for Theoretical Physics, National Academy of Sciences of Ukraine, Kiev, Ukraine\\
$^{4}$ Bose Institute, Department of Physics  and Centre for Astroparticle Physics and Space Science (CAPSS), Kolkata, India\\
$^{5}$ California Polytechnic State University, San Luis Obispo, California, United States\\
$^{6}$ Central China Normal University, Wuhan, China\\
$^{7}$ Centro de Aplicaciones Tecnol\'{o}gicas y Desarrollo Nuclear (CEADEN), Havana, Cuba\\
$^{8}$ Centro de Investigaci\'{o}n y de Estudios Avanzados (CINVESTAV), Mexico City and M\'{e}rida, Mexico\\
$^{9}$ Chicago State University, Chicago, Illinois, United States\\
$^{10}$ China Institute of Atomic Energy, Beijing, China\\
$^{11}$ China University of Geosciences, Wuhan, China\\
$^{12}$ Chungbuk National University, Cheongju, Republic of Korea\\
$^{13}$ Comenius University Bratislava, Faculty of Mathematics, Physics and Informatics, Bratislava, Slovak Republic\\
$^{14}$ COMSATS University Islamabad, Islamabad, Pakistan\\
$^{15}$ Creighton University, Omaha, Nebraska, United States\\
$^{16}$ Department of Physics, Aligarh Muslim University, Aligarh, India\\
$^{17}$ Department of Physics, Pusan National University, Pusan, Republic of Korea\\
$^{18}$ Department of Physics, Sejong University, Seoul, Republic of Korea\\
$^{19}$ Department of Physics, University of California, Berkeley, California, United States\\
$^{20}$ Department of Physics, University of Oslo, Oslo, Norway\\
$^{21}$ Department of Physics and Technology, University of Bergen, Bergen, Norway\\
$^{22}$ Dipartimento di Fisica, Universit\`{a} di Pavia, Pavia, Italy\\
$^{23}$ Dipartimento di Fisica dell'Universit\`{a} and Sezione INFN, Cagliari, Italy\\
$^{24}$ Dipartimento di Fisica dell'Universit\`{a} and Sezione INFN, Trieste, Italy\\
$^{25}$ Dipartimento di Fisica dell'Universit\`{a} and Sezione INFN, Turin, Italy\\
$^{26}$ Dipartimento di Fisica e Astronomia dell'Universit\`{a} and Sezione INFN, Bologna, Italy\\
$^{27}$ Dipartimento di Fisica e Astronomia dell'Universit\`{a} and Sezione INFN, Catania, Italy\\
$^{28}$ Dipartimento di Fisica e Astronomia dell'Universit\`{a} and Sezione INFN, Padova, Italy\\
$^{29}$ Dipartimento di Fisica `E.R.~Caianiello' dell'Universit\`{a} and Gruppo Collegato INFN, Salerno, Italy\\
$^{30}$ Dipartimento DISAT del Politecnico and Sezione INFN, Turin, Italy\\
$^{31}$ Dipartimento di Scienze MIFT, Universit\`{a} di Messina, Messina, Italy\\
$^{32}$ Dipartimento Interateneo di Fisica `M.~Merlin' and Sezione INFN, Bari, Italy\\
$^{33}$ European Organization for Nuclear Research (CERN), Geneva, Switzerland\\
$^{34}$ Faculty of Electrical Engineering, Mechanical Engineering and Naval Architecture, University of Split, Split, Croatia\\
$^{35}$ Faculty of Engineering and Science, Western Norway University of Applied Sciences, Bergen, Norway\\
$^{36}$ Faculty of Nuclear Sciences and Physical Engineering, Czech Technical University in Prague, Prague, Czech Republic\\
$^{37}$ Faculty of Physics, Sofia University, Sofia, Bulgaria\\
$^{38}$ Faculty of Science, P.J.~\v{S}af\'{a}rik University, Ko\v{s}ice, Slovak Republic\\
$^{39}$ Frankfurt Institute for Advanced Studies, Johann Wolfgang Goethe-Universit\"{a}t Frankfurt, Frankfurt, Germany\\
$^{40}$ Fudan University, Shanghai, China\\
$^{41}$ Gangneung-Wonju National University, Gangneung, Republic of Korea\\
$^{42}$ Gauhati University, Department of Physics, Guwahati, India\\
$^{43}$ Helmholtz-Institut f\"{u}r Strahlen- und Kernphysik, Rheinische Friedrich-Wilhelms-Universit\"{a}t Bonn, Bonn, Germany\\
$^{44}$ Helsinki Institute of Physics (HIP), Helsinki, Finland\\
$^{45}$ High Energy Physics Group,  Universidad Aut\'{o}noma de Puebla, Puebla, Mexico\\
$^{46}$ Horia Hulubei National Institute of Physics and Nuclear Engineering, Bucharest, Romania\\
$^{47}$ Indian Institute of Technology Bombay (IIT), Mumbai, India\\
$^{48}$ Indian Institute of Technology Indore, Indore, India\\
$^{49}$ INFN, Laboratori Nazionali di Frascati, Frascati, Italy\\
$^{50}$ INFN, Sezione di Bari, Bari, Italy\\
$^{51}$ INFN, Sezione di Bologna, Bologna, Italy\\
$^{52}$ INFN, Sezione di Cagliari, Cagliari, Italy\\
$^{53}$ INFN, Sezione di Catania, Catania, Italy\\
$^{54}$ INFN, Sezione di Padova, Padova, Italy\\
$^{55}$ INFN, Sezione di Pavia, Pavia, Italy\\
$^{56}$ INFN, Sezione di Torino, Turin, Italy\\
$^{57}$ INFN, Sezione di Trieste, Trieste, Italy\\
$^{58}$ Inha University, Incheon, Republic of Korea\\
$^{59}$ Institute for Gravitational and Subatomic Physics (GRASP), Utrecht University/Nikhef, Utrecht, Netherlands\\
$^{60}$ Institute of Experimental Physics, Slovak Academy of Sciences, Ko\v{s}ice, Slovak Republic\\
$^{61}$ Institute of Physics, Homi Bhabha National Institute, Bhubaneswar, India\\
$^{62}$ Institute of Physics of the Czech Academy of Sciences, Prague, Czech Republic\\
$^{63}$ Institute of Space Science (ISS), Bucharest, Romania\\
$^{64}$ Institut f\"{u}r Kernphysik, Johann Wolfgang Goethe-Universit\"{a}t Frankfurt, Frankfurt, Germany\\
$^{65}$ Instituto de Ciencias Nucleares, Universidad Nacional Aut\'{o}noma de M\'{e}xico, Mexico City, Mexico\\
$^{66}$ Instituto de F\'{i}sica, Universidade Federal do Rio Grande do Sul (UFRGS), Porto Alegre, Brazil\\
$^{67}$ Instituto de F\'{\i}sica, Universidad Nacional Aut\'{o}noma de M\'{e}xico, Mexico City, Mexico\\
$^{68}$ iThemba LABS, National Research Foundation, Somerset West, South Africa\\
$^{69}$ Jeonbuk National University, Jeonju, Republic of Korea\\
$^{70}$ Johann-Wolfgang-Goethe Universit\"{a}t Frankfurt Institut f\"{u}r Informatik, Fachbereich Informatik und Mathematik, Frankfurt, Germany\\
$^{71}$ Korea Institute of Science and Technology Information, Daejeon, Republic of Korea\\
$^{72}$ KTO Karatay University, Konya, Turkey\\
$^{73}$ Laboratoire de Physique des 2 Infinis, Ir\`{e}ne Joliot-Curie, Orsay, France\\
$^{74}$ Laboratoire de Physique Subatomique et de Cosmologie, Universit\'{e} Grenoble-Alpes, CNRS-IN2P3, Grenoble, France\\
$^{75}$ Lawrence Berkeley National Laboratory, Berkeley, California, United States\\
$^{76}$ Lund University Department of Physics, Division of Particle Physics, Lund, Sweden\\
$^{77}$ Nagasaki Institute of Applied Science, Nagasaki, Japan\\
$^{78}$ Nara Women{'}s University (NWU), Nara, Japan\\
$^{79}$ National and Kapodistrian University of Athens, School of Science, Department of Physics , Athens, Greece\\
$^{80}$ National Centre for Nuclear Research, Warsaw, Poland\\
$^{81}$ National Institute of Science Education and Research, Homi Bhabha National Institute, Jatni, India\\
$^{82}$ National Nuclear Research Center, Baku, Azerbaijan\\
$^{83}$ National Research and Innovation Agency - BRIN, Jakarta, Indonesia\\
$^{84}$ Niels Bohr Institute, University of Copenhagen, Copenhagen, Denmark\\
$^{85}$ Nikhef, National institute for subatomic physics, Amsterdam, Netherlands\\
$^{86}$ Nuclear Physics Group, STFC Daresbury Laboratory, Daresbury, United Kingdom\\
$^{87}$ Nuclear Physics Institute of the Czech Academy of Sciences, Husinec-\v{R}e\v{z}, Czech Republic\\
$^{88}$ Oak Ridge National Laboratory, Oak Ridge, Tennessee, United States\\
$^{89}$ Ohio State University, Columbus, Ohio, United States\\
$^{90}$ Physics department, Faculty of science, University of Zagreb, Zagreb, Croatia\\
$^{91}$ Physics Department, Panjab University, Chandigarh, India\\
$^{92}$ Physics Department, University of Jammu, Jammu, India\\
$^{93}$ Physics Program and International Institute for Sustainability with Knotted Chiral Meta Matter (SKCM2), Hiroshima University, Hiroshima, Japan\\
$^{94}$ Physikalisches Institut, Eberhard-Karls-Universit\"{a}t T\"{u}bingen, T\"{u}bingen, Germany\\
$^{95}$ Physikalisches Institut, Ruprecht-Karls-Universit\"{a}t Heidelberg, Heidelberg, Germany\\
$^{96}$ Physik Department, Technische Universit\"{a}t M\"{u}nchen, Munich, Germany\\
$^{97}$ Politecnico di Bari and Sezione INFN, Bari, Italy\\
$^{98}$ Research Division and ExtreMe Matter Institute EMMI, GSI Helmholtzzentrum f\"ur Schwerionenforschung GmbH, Darmstadt, Germany\\
$^{99}$ Saga University, Saga, Japan\\
$^{100}$ Saha Institute of Nuclear Physics, Homi Bhabha National Institute, Kolkata, India\\
$^{101}$ School of Physics and Astronomy, University of Birmingham, Birmingham, United Kingdom\\
$^{102}$ Secci\'{o}n F\'{\i}sica, Departamento de Ciencias, Pontificia Universidad Cat\'{o}lica del Per\'{u}, Lima, Peru\\
$^{103}$ Stefan Meyer Institut f\"{u}r Subatomare Physik (SMI), Vienna, Austria\\
$^{104}$ SUBATECH, IMT Atlantique, Nantes Universit\'{e}, CNRS-IN2P3, Nantes, France\\
$^{105}$ Sungkyunkwan University, Suwon City, Republic of Korea\\
$^{106}$ Suranaree University of Technology, Nakhon Ratchasima, Thailand\\
$^{107}$ Technical University of Ko\v{s}ice, Ko\v{s}ice, Slovak Republic\\
$^{108}$ The Henryk Niewodniczanski Institute of Nuclear Physics, Polish Academy of Sciences, Cracow, Poland\\
$^{109}$ The University of Texas at Austin, Austin, Texas, United States\\
$^{110}$ Universidad Aut\'{o}noma de Sinaloa, Culiac\'{a}n, Mexico\\
$^{111}$ Universidade de S\~{a}o Paulo (USP), S\~{a}o Paulo, Brazil\\
$^{112}$ Universidade Estadual de Campinas (UNICAMP), Campinas, Brazil\\
$^{113}$ Universidade Federal do ABC, Santo Andre, Brazil\\
$^{114}$ University of Cape Town, Cape Town, South Africa\\
$^{115}$ University of Houston, Houston, Texas, United States\\
$^{116}$ University of Jyv\"{a}skyl\"{a}, Jyv\"{a}skyl\"{a}, Finland\\
$^{117}$ University of Kansas, Lawrence, Kansas, United States\\
$^{118}$ University of Liverpool, Liverpool, United Kingdom\\
$^{119}$ University of Science and Technology of China, Hefei, China\\
$^{120}$ University of South-Eastern Norway, Kongsberg, Norway\\
$^{121}$ University of Tennessee, Knoxville, Tennessee, United States\\
$^{122}$ University of the Witwatersrand, Johannesburg, South Africa\\
$^{123}$ University of Tokyo, Tokyo, Japan\\
$^{124}$ University of Tsukuba, Tsukuba, Japan\\
$^{125}$ University Politehnica of Bucharest, Bucharest, Romania\\
$^{126}$ Universit\'{e} Clermont Auvergne, CNRS/IN2P3, LPC, Clermont-Ferrand, France\\
$^{127}$ Universit\'{e} de Lyon, CNRS/IN2P3, Institut de Physique des 2 Infinis de Lyon, Lyon, France\\
$^{128}$ Universit\'{e} de Strasbourg, CNRS, IPHC UMR 7178, F-67000 Strasbourg, France, Strasbourg, France\\
$^{129}$ Universit\'{e} Paris-Saclay Centre d'Etudes de Saclay (CEA), IRFU, D\'{e}partment de Physique Nucl\'{e}aire (DPhN), Saclay, France\\
$^{130}$ Universit\`{a} degli Studi di Foggia, Foggia, Italy\\
$^{131}$ Universit\`{a} del Piemonte Orientale, Vercelli, Italy\\
$^{132}$ Universit\`{a} di Brescia, Brescia, Italy\\
$^{133}$ Variable Energy Cyclotron Centre, Homi Bhabha National Institute, Kolkata, India\\
$^{134}$ Warsaw University of Technology, Warsaw, Poland\\
$^{135}$ Wayne State University, Detroit, Michigan, United States\\
$^{136}$ Westf\"{a}lische Wilhelms-Universit\"{a}t M\"{u}nster, Institut f\"{u}r Kernphysik, M\"{u}nster, Germany\\
$^{137}$ Wigner Research Centre for Physics, Budapest, Hungary\\
$^{138}$ Yale University, New Haven, Connecticut, United States\\
$^{139}$ Yonsei University, Seoul, Republic of Korea\\
$^{140}$  Zentrum  f\"{u}r Technologie und Transfer (ZTT), Worms, Germany\\
$^{141}$ Affiliated with an institute covered by a cooperation agreement with CERN\\
$^{142}$ Affiliated with an international laboratory covered by a cooperation agreement with CERN.\\

\end{flushleft} 

\end{document}